\documentclass[english,journal=jctcce,manuscript=article,articletitle=true,layout=twocolumn]{achemso}
\usepackage[T1]{fontenc}
\usepackage[utf8]{inputenc}
\usepackage{babel}
\usepackage{prettyref}
\usepackage{textcomp}
\usepackage{amsmath}
\usepackage{amsthm}
\usepackage{amssymb}
\usepackage{graphicx}
\usepackage[numbers]{natbib}
\usepackage[unicode=true,pdfusetitle,
 bookmarks=true,bookmarksnumbered=false,bookmarksopen=false,
 breaklinks=false,pdfborder={0 0 1},backref=false,colorlinks=false]
 {hyperref}

\makeatletter

\title{Automatic Generation of Accurate and Cost-efficient Auxiliary Basis
Sets}

\author{Susi Lehtola}

\email{susi.lehtola@alumni.helsinki.fi}

\affiliation{Molecular Sciences Software Institute, Blacksburg, Virginia 24061,
United States}

\alsoaffiliation{Department of Chemistry, University of Helsinki, P.O. Box 55, FI-00014
University of Helsinki, Finland}

\providecommand{\tabularnewline}{\\}
\newcommand{\lyxdot}{.}

\usepackage[version=4]{mhchem}

\usepackage{cleveref}
\AtBeginDocument{%
\let\ref\cref
}

\@ifundefined{showcaptionsetup}{}{%
 \PassOptionsToPackage{caption=false}{subfig}}
\usepackage{subfig}
\makeatother

\begin{document}
\begin{abstract}
We have recently discussed an algorithm to automatically generate
auxiliary basis sets (ABSs) of the standard form for density fitting
(DF) or resolution-of-the-identity (RI) calculations in a given atomic
orbital basis set (OBS) of any form, such as Gaussian-type orbitals,
Slater-type orbitals, or numerical atomic orbitals {[}J. Chem. Theory
Comput. 2021, 17, 6886{]}. In this work, we study two ways to reduce
the cost of such automatically generated ABSs without sacrificing
their accuracy. We contract the ABS with a singular value decomposition
proposed by Kállay {[}J. Chem. Phys. 2014, 141, 244113{]}, used here
in a somewhat different setting. We also drop the high-angular momentum
functions from the ABS, as they are unnecessary for global fitting
methods. Studying the effect of these two types of truncations on
Hartree--Fock (HF) and second-order Møller--Plesset perturbation
theory (MP2) calculations on a chemically diverse set of first- and
second-row molecules within the RI/DF approach, we show that accurate
total and atomization energies can be achieved by a combination of
the two approaches with significant reductions in the size of the
ABS. While the original approach yields ABSs whose number of functions
$N_{\text{bf}}^{\text{ABS}}$ scales with the number of functions
in the OBS, $N_{\text{OBS}}^{\text{bf}}$, as $N_{\text{ABS}}^{\text{bf}}=\gamma N_{\text{OBS}}^{\text{bf}}$
with the prefactor $\gamma\approx\mathcal{O}(10)$, the reduction
schemes of this work afford results of essentially the same quality
as the original unpruned and uncontracted ABS with $\gamma\approx5\text{–}6$,
while an accuracy that may suffice for routine applications is achievable
with a further reduced ABS with $\gamma\approx3\text{–}4$. The
observed errors are similar at HF and MP2 levels of theory, suggesting
that the generated ABSs are highly transferable, and can also be applied
to model challenging properties with high-level methods.
\end{abstract}
\newcommand*\ie{{\em i.e.}}
\newcommand*\eg{{\em e.g.}}
\newcommand*\etal{{\em et al.}}
\newcommand*\citeref[1]{ref. \citenum{#1}}
\newcommand*\citerefs[1]{refs. \citenum{#1}} 

\newcommand*\Erkale{{\sc Erkale}}
\newcommand*\Bagel{{\sc Bagel}}
\newcommand*\FHIaims{{\sc FHI-aims}}
\newcommand*\LibXC{{\sc LibXC}}
\newcommand*\Orca{{\sc Orca}}
\newcommand*\PySCF{{\sc PySCF}}
\newcommand*\PsiFour{{\sc Psi4}}
\newcommand*\Turbomole{{\sc Turbomole}}

\section{Introduction \label{sec:Introduction}}

The density fitting\citep{Whitten1973_JCP_4496,Baerends1973_CP_41,Dunlap1977_IJQC_81,Dunlap1979_JCP_3396,Dunlap2010_MP_3167}
(DF) also known as the the resolution of the identity\citep{Vahtras1993_CPL_514}
(RI) technique has become one of the mainstays of quantum chemistry.
The DF/RI approach is widely available in many program packages as
an optional technique to accelerate self-consistent field (SCF) as
well as post-Hartree--Fock calculations. It is used by default in
some packages, such as Psi4\citep{Smith2020_JCP_184108} and Orca\citep{Neese2022_WCMS_1606},
while other packages like BAGEL\citep{Shiozaki2018_WIRCMS_1331} and
Demon2K\citep{Geudtner2012_WIRCMS_548} go even further by relying
solely on the use of DF/RI for all calculations, thus foregoing traditional
routines based on the exact electron repulsion integrals (ERIs).

Assuming that the basis-functions are real-valued, these ERIs are
typically written in the Mulliken notation as 
\begin{equation}
(\mu\nu|\sigma\rho)=\int\frac{\chi_{\mu}(\boldsymbol{r})\chi_{\nu}(\boldsymbol{r})\chi_{\sigma}(\boldsymbol{r}')\chi_{\rho}(\boldsymbol{r}')}{|\boldsymbol{r}-\boldsymbol{r}'|}{\rm d}^{3}r{\rm d}^{3}r'\label{eq:tei}
\end{equation}
where $\mu$, $\nu$, $\sigma$, and $\rho$ are indices of atomic-orbital
basis functions ${\chi_{\mu}}$ that together form the orbital basis
set (OBS). In the DF/RI approach, the two-electron integrals of the
OBS are approximated with the help of an auxiliary basis set (ABS)
as\citep{Vahtras1993_CPL_514}
\begin{equation}
(\mu\nu|\sigma\rho)\approx\sum_{AB}(\mu\nu|A)(A|B)^{-1}(B|\sigma\rho),\label{eq:RI}
\end{equation}
where $A$ and $B$ are functions in the ABS and $(A|B)^{-1}$ denotes
the $A,B$ element of the inverse of the two-index Coulomb overlap
matrix $(A|B)$. Traditionally, ABSs are manually optimized for ground-state
energy calculations for each OBS for each level of theory,\citep{Hill2013_IJQC_21,Pedersen2023__}
but some automatic algorithms for generating an ABS from the given
OBS have also been suggested;\citep{Aquilante2007_JCP_114107,Yang2007_JCP_74102,Aquilante2009_JCP_154107,Ren2012_NJP_53020,Stoychev2017_JCTC_554,Lehtola2021_JCTC_6886}
see \citeref{Pedersen2023__} for a recent review.

While some of these automated schemes inherently assume the use of
Gaussian basis sets,\citep{Aquilante2009_JCP_154107,Yang2007_JCP_74102,Stoychev2017_JCTC_554}
we recently suggested an automated algorithm which can be used with
any type of atomic-orbital basis set. In short, the central idea of
this scheme is to generate the ABS by choosing it such that it spans
all one-center products $\chi_{\mu}\chi_{\nu}$ in the OBS to a given
degree of precision.

Because the angular part is closed---the product of two spherical
harmonics yielding a linear combination of spherical harmonics---the
task can be reduced to studying radial functions that should be well-described
by the sought-for ABS. As the one-center OBS products $\chi_{\mu}\chi_{\nu}$
can couple to angular momentum $|l_{\mu}-l_{\nu}|\leq L\leq l_{\mu}+l_{\nu}$,
where $l_{\mu}$ and $l_{\nu}$ are the angular momenta of the functions
$\chi_{\mu}$ and $\chi_{\nu}$, respectively, the algorithm of \citeref{Lehtola2021_JCTC_6886}
works by assembling a list of normalized product radial functions
\begin{equation}
R_{I}^{L}(r)=N_{I}\chi_{\mu_{I}}(r)\chi_{\nu_{I}}(r)\label{eq:candidate-func}
\end{equation}
for all possible angular momenta $L$ in the auxiliary basis set,
$N_{I}$ being the normalization coefficient in \ref{eq:candidate-func}.
These product functions $R_{I}^{L}$ serve as candidate auxiliary
basis functions for each shell $L$. The set of candidate radial functions
can be prescreened by a pivoted Cholesky decomposition of the ERI
tensor;\citep{Beebe1977_IJQC_683} this yields a more compact auxiliary
basis set at no loss in accuracy.\citep{Lehtola2021_JCTC_6886}

Next, the algorithm chooses the set of auxiliary radial functions
for each angular momentum $L$, $R_{I}^{L}(r)$, by a pivoted Cholesky
decomposition of the one-center Coulomb overlap matrix $(R_{I}^{L}|R_{J}^{L})$,
motivated by earlier work of \citet{Aquilante2009_JCP_154107} The
decomposition is carried out to the given threshold $\tau$. The resulting
set of functions is capable of reproducing all the functions in the
candidates pool $\{R_{I}^{L}\}$ to within the used threshold $\tau$,
thus affording an optimal\citep{Harbrecht2012_ANM_428} black-box
method to generate auxiliary basis sets.

The algorithm of \citeref{Lehtola2021_JCTC_6886} is applicable to
any atomic-orbital basis sets: the necessary equations to compute
the one-center Coulomb overlap matrix for Gaussian-type orbitals (GTOs)
as well as Slater-type orbitals (STOs) were given in \citeref{Lehtola2021_JCTC_6886},
while the necessary integrals for numerical atomic orbitals\citep{Lehtola2019_IJQC_25968}
(NAOs) can easily be computed by quadrature.\citep{Lehtola2019_IJQC_25945} 

As a side note, we have also successfully used an analogous technique
to cure OBS overcompleteness in SCF calculations by solving the Roothaan
equations\citep{Roothaan1951_RMP_69} $\boldsymbol{F}\boldsymbol{C}=\boldsymbol{S}\boldsymbol{C}\boldsymbol{E}$,
where $\boldsymbol{F}$ is the Fock matrix, $\boldsymbol{S}$ is the
overlap matrix, and\textbf{ }$\boldsymbol{C}$ and $\boldsymbol{E}$
are the matrix of orbital expansion coefficients and the corresponding
diagonal matrix of orbital energies, in a subbasis chosen by the pivoted
Cholesky decomposition of $\boldsymbol{S}$.\citep{Lehtola2019_JCP_241102,Lehtola2020_PRA_32504}
Choosing the basis functions by a pivoted Cholesky decomposition of
the (Coulomb) overlap matrix---which is a so-called Gram matrix---is
analogous to using an optimal Gram--Schmidt orthogonalization.\citep{Pedersen2023__}

Even if automatically generated ABSs tend to be large compared to
their possible manually optimized counterparts, they can be extremely
useful for practical applications. Although an automatically generated
auxiliary basis set may be several times larger than a manually optimized
conterpart, the cost of DF/RI methods scales only linearly with the
size of the ABS. The automatic auxiliary basis set machinery enables
the use of the DF/RI technique in all OBSs, offering significant speedups
in calculations compared to the use of the exact TEIs, because the
factorization of \ref{eq:RI} can be employed to formulate efficient
implementations of many quantum chemical models. Therefore, DF/RI
calculations with automatically generated basis sets tend to not only
be considerably faster than calculations with exact ERIs, but also
not orders of magnitude slower than DF/RI calculations employing hand-optimized
ABSs---provided that such ABSs exist for the employed OBS and the
studied property at the considered level of theory.

Our previous study\citep{Lehtola2021_JCTC_6886} focused on showing
that negligible errors in Hartree--Fock (HF) and second-order Møller--Plesset
perturbation theory (MP2) total energies could be achieved with the
technique proposed therein. The calculations in \citeref{Lehtola2021_JCTC_6886}
were carried out with GTO basis sets to allow the straightforward
computation of exact values without the DF/RI approximation.

In this work, we will focus on improving the cost-efficiency of the
automatically generated ABS. We will accomplish this along two routes.
First, we will consider contracting the ABS employing a singular value
decomposition (SVD). The use of this technique to reduce the cost
of RI/DF methods was originally suggested by \citet{Kallay2014_JCP_244113}
within a different context: system-specific truncations of fixed ABSs
in polyatomic calculations. In this work, the contraction of the ABS
serves to restore the energetic connection of the autogenerated basis
to the two-electron integrals, which was severed in the algorithm
of \citeref{Lehtola2021_JCTC_6886} that was laid out above. The contraction
procedure is especially useful in the case of contracted Gaussian
OBSs, as the resulting contracted ABSs turn out to have fewer auxiliary
basis functions. The contraction procedure is also seen to be beneficial
with uncontracted Gaussian basis sets, as it results in a reduced
number of auxiliary basis functions. We expect the contraction procedure
to be useful especially for NAOs, because the recontraction of NAOs
can be carried out at no cost to integral evaluation.

Second, we will also consider two schemes to discard high-angular
momentum (HAM) functions from the ABS: in addition to the scheme used
by previous automated schemes (\citerefs{Yang2007_JCP_74102} and
\citenum{Stoychev2017_JCTC_554}), we will also discuss a first-principles
scheme, which has been previously discussed in the context of optimized
ABSs.\citep{Weigend2008_JCC_167} As we will demonstrate in this work,
the HAM functions can be safely discarded in global fitting methods,
as they have negligible effect on total and relative energies.

The layout of this work is the following. Next, in \ref{sec:Theory},
we will discuss the theory behind the present approach: the contraction
of the ABS is discussed in \ref{subsec:Contracted-sets}, while the
pruning of HAM functions is discussed in \ref{subsec:Reduced-angular-momentum}.
The computational details of the benchmarking based on HF and MP2
calculations are given in \ref{sec:Computational-Details}. The results
of the study are discussed in \ref{sec:Results}, and the study concludes
in a brief summary and discussion in \ref{sec:Summary-and-Discussion}.
Atomic units are used throughout, unless specified otherwise.

\section{Theory \label{sec:Theory}}

\subsection{Contracted sets \label{subsec:Contracted-sets}}

For clarity, we will reuse the notation of \citet{Kallay2014_JCP_244113},
who writes \ref{eq:RI} in the form
\begin{equation}
\boldsymbol{\mathcal{I}}\approx\boldsymbol{I}\boldsymbol{V}^{-1}\boldsymbol{I}^{\text{T}}\label{eq:ri-kallay}
\end{equation}
where $\mathcal{I}_{\mu\nu,\rho\sigma}=(\mu\nu|\rho\sigma)$ is the
ERI tensor, $I_{\mu\nu,A}=(\mu\nu|A)$ are the three-index integrals
and $V_{A,B}=(A|B)$ are the two-index Coulomb overlap integrals.
\Cref{eq:ri-kallay} can be expressed in a more compact form in the
orthogonalized ABS: defining orthogonalized three-index integrals,
\begin{equation}
\boldsymbol{J}=\boldsymbol{I}\boldsymbol{V}^{-1/2}\label{eq:J}
\end{equation}
\ref{eq:ri-kallay} is given simply by 
\begin{equation}
\boldsymbol{\mathcal{I}}\approx\boldsymbol{J}\boldsymbol{J}^{\text{T}}.\label{eq:ri-kallay-short}
\end{equation}

\citet{Kallay2014_JCP_244113} pointed out that the size of the auxiliary
basis can be reduced by a SVD of $\boldsymbol{J}$ by throwing out
vectors with suitably small singular values. The singular values can
be computed as the eigenvalues of the symmetric matrix
\begin{equation}
\boldsymbol{W}=\boldsymbol{J}^{\text{T}}\boldsymbol{J}\label{eq:W}
\end{equation}
which has the dimension $N_{\text{aux}}\times N_{\text{aux}}$ where
$N_{\text{aux}}$ is the number of auxiliary basis functions. Because
the number of auxiliary functions typically scales as the number of
orbital basis functions, the matrix \textbf{$\boldsymbol{W}$} of
\ref{eq:W} can be diagonalized even in large calculations.

In this work, we use the above method to contract the atomic auxiliary
basis. This merely requires forming $\boldsymbol{W}$ for each element,
choosing the contraction based on the singular values, and repeating
the procedure for all elements in the basis set.

Because the $\boldsymbol{W}$ matrix defined by \ref{eq:W} is obtained
by summing over all OBS products on the element, $\boldsymbol{W}$
turns out to have a special angular structure. The only significant
values are found when the angular momenta of the bra and ket basis
functions coincide, $l'=l$ and $m'=m$, the other elements of $\boldsymbol{W}$
being numerically zero. As one could expect, all spatial directions
are therefore averaged in $\boldsymbol{W}$.

Because of this special structure, it suffices to compute, e.g., the
$m'=0=m$ subblock of the $l=l'$ angular submatrix of $\boldsymbol{W}$,
which we denote as $\boldsymbol{L}$, which contains information on
the importance of the radial auxiliary functions of angular momentum
$l$.

The eigendecomposition of $\boldsymbol{L}$ then yields
\begin{equation}
\boldsymbol{L}=\boldsymbol{U}\boldsymbol{\Lambda}\boldsymbol{U}^{\text{T}},\label{eq:Weig}
\end{equation}
where $\boldsymbol{\Lambda}$ is a diagonal matrix of eigenvalues
and $\boldsymbol{U}$ is the corresponding matrix of eigenvectors.
Now, we can choose a reduced ABS corresponding to some precision $\epsilon$
by including only those columns $\boldsymbol{U}_{i}$ that correspond
to eigenvalues $\lambda_{i}\ge\epsilon$. Such an ABS has guaranteed
accuracy, as the three-index integrals are reproduced to the precision
specified by $\epsilon$.\citep{Kallay2014_JCP_244113}

The contraction algorithm is thus the following:
\begin{enumerate}
\item Define the contraction threshold $\epsilon$ and read in the OBS and
the ABS.
\item Loop over atoms, and the atomic angular momentum $l$ in the ABS
\begin{enumerate}
\item Form the $l=l'$ and $m=0=m'$ angular subblock of $\boldsymbol{W}$
from \ref{eq:W}, denoted as $\boldsymbol{L}$.
\item Compute the eigenvectors $\boldsymbol{U}_{i}$ with eigenvalues $\lambda_{i}$
of $\boldsymbol{L}$.
\item Include the eigenvectors $\boldsymbol{U}_{i}$ in the auxiliary basis
if their eigenvalues satisfy $\lambda_{i}\ge\epsilon$ .
\item Convert the eigenvectors into the original non-orthonormal basis by
$\boldsymbol{C}=\boldsymbol{V}^{-1/2}\boldsymbol{U}$, where $\boldsymbol{V}^{-1/2}$
is the orthogonalizing matrix of \ref{eq:J}.
\end{enumerate}
\end{enumerate}
To extract the contraction coefficients of a GTO auxiliary basis,
a further step is necessary. As Gaussian basis functions are by convention
tabulated in terms of primitives normalized in the overlap metric,
while the coefficients of the eigenvectors $\boldsymbol{C}$ formed
above correspond to Coulomb normalized primitives, one has to convert
the contraction coefficients into the expected normalization. This
is achieved by scaling the coefficient of the $\mu$:th Gaussian primitive
in the $i$:th auxiliary function, $C_{\mu i}$, by the square root
of the corresponding Gaussian exponent, $C_{\mu i}\to C_{\mu i}\sqrt{\alpha_{\mu}}$,
see the Appendix for a derivation.

\subsection{Pruned Angular Momentum \label{subsec:Reduced-angular-momentum}}

By default, automatically generated basis sets arising from a large
orbital basis set may contain functions of extremely HAM, as an orbital
basis with maximum angular momentum $l_{\text{OBS}}^{\text{max}}$
will give rise to orbital products with maximum angular momentum $2l_{\text{OBS}}^{\text{max}}$.
For instance, a polarized quintuple-$\zeta$ basis for the oxygen
atom ($l_{\text{OBS}}^{\text{max}}=5$) will lead to an auxiliary
basis set with up to $l=10$ functions.

Such functions are problematic in Gaussian-basis programs, as most
programs do not implement integrals to such HAM. Moreover, since Gaussian
integrals are typically computed in the cartesian Gaussian basis,
computing integrals with HAM functions is extremely costly: the number
of cartesian functions of angular momentum $l$ is $(l+1)(l+2)/2$
compared to $2l+1$ for spherical functions. Furthermore, the HAM
functions' contributions to total ground state energies tend to be
negligible. This motivates removing them from the ABS altogether by
a pruning procedure.

The question of what HAM functions can be removed is best approached
by asking the opposite question: what are the functions which should
never be pruned from the ABS? 

Even in highly excited states, most of the total energy arises from
orbitals that are occupied in the atomic ground state. The correct
description of occupied orbitals being of key importance, the pruned
auxiliary basis set should therefore reproduce (i) the highly energetic
interactions of the atom's occupied orbitals with each other, as well
as (ii) the interactions of the atom's occupied orbitals with its
unoccupied orbitals, which are important for polarization and correlation
effects in polyatomic systems.

Criterion (i) is automatically satisfied, if all products of occupied
orbitals are correctly reproduced by the auxiliary basis set. Therefore,
we demand that the maximum angular momentum of the functions kept
in the auxiliary basis, $l_{\text{keep}}^{\text{max}}$, is at least
two times the maximum angular momentum of occupied shells in the ground
state of the atom $l_{\text{occ}}^{\text{max}}$, $l_{\text{keep}}^{\text{max}}\ge2l_{\text{occ}}^{\text{max}}$. 

\citet{Yang2007_JCP_74102} define $l_{\text{occ}}^{\text{max}}$
as having the values 0, 1, 2, and 3 for $Z\leq2$, $3\leq Z\leq18$,
$19\leq Z\leq54$, and $Z\ge55$, respectively, while \citet{Stoychev2017_JCTC_554}
use the ranges $Z\leq2$, $3\leq Z\leq20$, $21\leq Z\leq56$, and
$Z\ge57$, respectively. That is, while both \citet{Yang2007_JCP_74102}
and \citet{Stoychev2017_JCTC_554} include $p$ functions for the
pre-$d$ Li and Be atoms, the former likewise add $d$ and $f$ functions
for the pre-$d$ and pre-$f$ alkali and alkaline atoms, respectively,
while the latter do not. We will follow \citet{Yang2007_JCP_74102}
and include $d$ and $f$ functions for K and Ca, and Cs and Ba, respectively.

Criterion (ii) for the minimal accuracy of the auxiliary basis set
is satisfied, if the auxiliary basis is able to accurately describe
products of an occupied atomic orbital with any other orbital. Denoting
the maximum angular momentum of the orbital basis by $l_{\text{OBS}}^{\text{max}}$,
the second condition is seen to be satisfied by $l_{\text{keep}}^{\text{max}}\ge l_{\text{occ}}^{\text{max}}+l_{\text{OBS}}^{\text{max}}$.

The two criteria must be satisfied simultaneously, which is why we
choose
\begin{equation}
l_{\text{keep}}^{\text{max}}=\max\ (2l_{\text{occ}}^{\text{max}},l_{\text{occ}}^{\text{max}}+l_{\text{OBS}}^{\text{max}}+l_{\text{inc}}),\label{eq:lmax}
\end{equation}
where $l_{\text{inc}}$ is an adjustable parameter. As far as we are
aware, this criterion has not been studied in the literature in the
context of automatic generation of auxiliary basis sets, although
a similar discussion can be found in the work of \citet{Weigend2008_JCC_167}
discussing the optimization of auxiliary basis sets. (The implementation
of the atomic Cholesky procedures of \citet{Aquilante2007_JCP_114107}
and \citet{Aquilante2009_JCP_154107} in the OpenMolcas program\citep{Manni2023_JCTC_}
does include optional pruning of the angular momentum to a maximum
defined by the auxiliary basis sets of \citet{Weigend2008_JCC_167};
however, this procedure does not appear to be fully documented.)

In contrast, the work of \citet{Yang2007_JCP_74102} on Coulomb fitting
limited the maximum angular momentum by
\begin{equation}
l_{\text{Yang}}^{\text{max}}=\max\ (2l_{\text{occ}}^{\text{max}},l_{\text{OBS}}^{\text{max}}+l_{\text{inc}})\label{eq:lmax-yang}
\end{equation}
and employed $l_{\text{inc}}=1$ for all elements, while \citet{Stoychev2017_JCTC_554}
similarly employed \ref{eq:lmax-yang} with $l_{\text{inc}}=1$ for
elements up to Ar and 2 for other elements. 

Our truncation scheme (\ref{eq:lmax}) is seen to have important differences
from that used by \citet{Yang2007_JCP_74102} and \citet{Stoychev2017_JCTC_554}
(\ref{eq:lmax-yang}): \ref{eq:lmax} includes the atomic angular
momentum and therefore naturally employs a larger angular momentum
cutoff for heavier elements. 

We will compare the two schemes for pruning the angular momentum defined
by \cref{eq:lmax,eq:lmax-yang} in \ref{subsec:3ZaPa-NR}, examining
various values for $l_{\text{inc}}$ and comparing the results to
ones obtained with the full, unpruned and uncontracted autogenerated
auxiliary basis.

\section{Computational Details \label{sec:Computational-Details}}

The computational details of this work are similar to those of \citeref{Lehtola2021_JCTC_6886}.
As in the previous work, we will examine the accuracy of the scheme
on the triple-$\zeta$ to quintuple-$\zeta$ nZaPa-NR basis sets of
\citet{Ranasinghe2013_JCP_144104}, for which optimized auxiliary
basis sets have not been reported in the literature. These OBSs were
obtained from the Basis Set Exchange,\citep{Pritchard2019_JCIM_4814}
in whose Python backend we previously implemented\citep{Lehtola2021_JCTC_6886}
the AutoAux procedure of \citet{Stoychev2017_JCTC_554}, which will
again be used as a state-of-the-art point of reference. In contrast
to \citeref{Lehtola2021_JCTC_6886}, where all ABSs were generated
for fully uncontracted OBSs, the ABSs used in this work are generated
for contracted OBSs.

The full primitive ABSs were generated with ERKALE\citep{Lehtola2012_JCC_1572,Lehtola2018__a}
following the procedure of \citeref{Lehtola2021_JCTC_6886}, employing
the threshold $\tau=10^{-7}$ for the pivoted Cholesky decomposition.
Unless specified otherwise, the orbital products were prescreened
with a pivoted Cholesky decomposition of the ERI tensor, as recommended
in \citeref{Lehtola2021_JCTC_6886}.

The contraction and reduction schemes presented above in \ref{sec:Theory}
were implemented in ERKALE\citep{Lehtola2012_JCC_1572,Lehtola2018__a}
in this work, and they are freely and openly available on GitHub.
The employed values for the contraction and reduction parameters,
$\epsilon$ and $l_{\text{inc}}$, respectively, are discussed in
\ref{sec:Results}.

Our previous work\citep{Lehtola2021_JCTC_6886} studied HF and MP2
total energies in the non-multireference part of the W4-17 test set\citep{Karton2017_JCC_2063}
and showed that their RI/DF errors can be made negligible by the use
of suitably large primitive ABSs. As the ABSs generated by the previously
suggested automated procedure are already known to be transferable,\citep{Lehtola2021_JCTC_6886}
and the contraction and reduction schemes of \ref{sec:Theory} are
based on mathematical principles, we will study here the chemically
diverse subset of first- and second-row molecules from the database
of \citet{Weigend2005_PCCP_305}, which has been previously used to
assess DF/RI auxiliary basis sets for Hartree--Fock.\citep{Weigend2008_JCC_167}
We limit the calculations to molecules containing at most nine atoms,
which suffices to demonstrate the accuracy of the reduced basis sets.

As the employed spin states for the molecules are not documented in
\citerefs{Weigend2008_JCC_167} and \citenum{Weigend2005_PCCP_305},
we decided to run HF and MP2 calculations for the  \ce{Al} (doublet), \ce{Be}, \ce{B} (doublet), \ce{Cl} (doublet), \ce{C} (triplet), \ce{F} (doublet), \ce{H} (doublet), \ce{Li} (doublet), \ce{Mg}, \ce{Na} (doublet), \ce{N} (quartet), \ce{O} (triplet), \ce{P} (quartet), \ce{Si} (triplet), and \ce{S} (triplet)
atoms, and the \ce{AlN} (triplet), \ce{BeS} (triplet), \ce{Cl2}, \ce{ClF}, \ce{CO}, \ce{F2}, \ce{H2}, \ce{HCl}, \ce{HF}, \ce{Li2}, \ce{LiCl}, \ce{LiF}, \ce{LiH}, \ce{MgF} (doublet), \ce{N2}, \ce{NaCl}, \ce{NaF}, \ce{NaH}, \ce{P2}, \ce{S2} (triplet),  \ce{BeH2}, \ce{CO2}, \ce{CS2}, \ce{H2O}, \ce{HCN}, \ce{HCP}, \ce{HNC}, \ce{HNO}, \ce{HSH}, \ce{Li2O}, \ce{LiSLi}, \ce{MgCl2}, \ce{MgF2}, \ce{MgH2}, \ce{Na2O} (triplet), \ce{Na2S}, \ce{OF2}, \ce{SF2}, \ce{SiO2}, \ce{SiS2},  \ce{AlCl3}, \ce{AlF3}, \ce{AlH3}, \ce{Be4}, \ce{BF3}, \ce{BH3}, \ce{C2H2}, \ce{CH2O}, \ce{ClF3}, \ce{H2O2}, \ce{HNO2}, \ce{HSSH}, \ce{Mg4}, \ce{N2H2}, \ce{N4}, \ce{Na3N} (triplet), \ce{Na3P} (triplet), \ce{NF3}, \ce{NH3}, \ce{PF3}, \ce{PH3}, \ce{PLi3},  \ce{Al2O3}, \ce{Al2S3}, \ce{CF4}, \ce{CH2O2}, \ce{CH3N}, \ce{CH4}, \ce{HNO3}, \ce{S5}, \ce{SF4}, \ce{SiCl4}, \ce{SiF4}, \ce{SiH4},  \ce{Be2F4}, \ce{Be2H4}, \ce{BH3CO}, \ce{C2H3N}, \ce{C2H4}, \ce{CH3OH}, \ce{H2CO3}, \ce{LiBH4}, \ce{N2H4}, \ce{NH4F}, \ce{PF5},  \ce{H2SO4}, \ce{SF6},  \ce{B2H6}, \ce{B4H4}, \ce{BH3NH3}, \ce{C2H6}, \ce{C4H4}, \ce{H3PO4}, \ce{Li4Cl4}, \ce{Li4H4}, \ce{Li8},  \ce{BeC2H6}, and \ce{BeF2O2H4}
molecules. Altogether, the assessment therefore includes 113 systems,
composed of 15 atoms and 98 molecules.

Conventional HF and MP2 calculations were performed for these molecules
with the Gaussian'16 program;\citep{Frisch2016__} modifications of
the basis set were disabled with \texttt{IOp(3/60=\textminus 1)}.
The RI/DF calculations were carried out with Psi4.\citep{Smith2020_JCP_184108}
The basis set linear dependence threshold was set to $10^{-7}$ in
both programs.

As our test systems only include first- and second-row molecules,
the same value of $l_{\text{inc}}$ is used for all elements in \ref{eq:lmax-yang}
in our test calculations; note that \citet{Yang2007_JCP_74102} and
\citet{Stoychev2017_JCTC_554} similarly use $l_{\text{inc}}=1$ for
all elements up to argon. 

\section{Results \label{sec:Results}}

\subsection{Detailed Analysis in the 3ZaPa-NR Basis Set\label{subsec:3ZaPa-NR}}

We begin by analyzing how RI/DF errors in HF and MP2 total and atomization
energies are affected by contracting the auxiliary basis set according
to the procedure of \prettyref{subsec:Contracted-sets}, when the
3ZaPa-NR OBS is used. Because the DF/RI error is an extensive quantity,
we will examine errors in total energies \emph{per electron}, and
errors in atomization energies \emph{per atom.}

\begin{figure}
\begin{centering}
\subfloat[Total energies]{\begin{centering}
\includegraphics[width=1\linewidth]{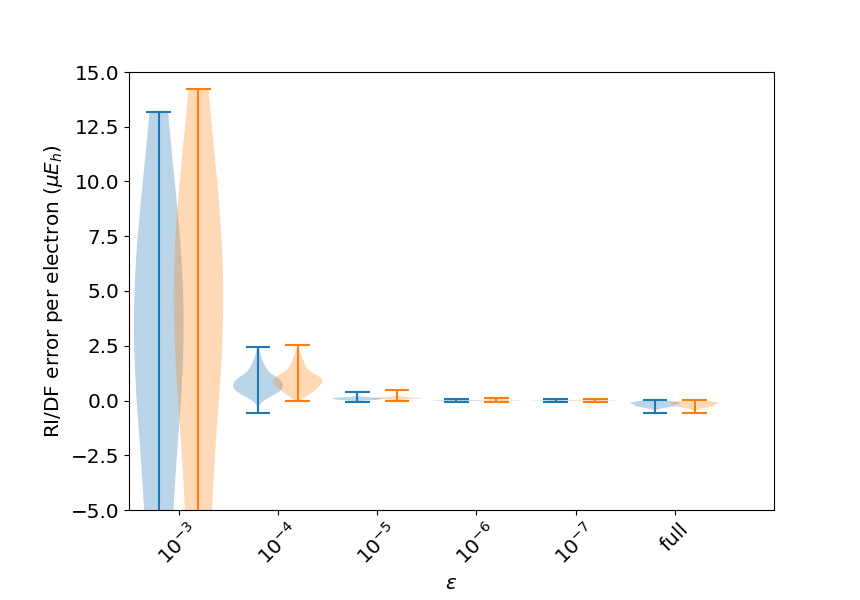}
\par\end{centering}
}
\par\end{centering}
\subfloat[Atomization energies]{\begin{centering}
\includegraphics[width=1\linewidth]{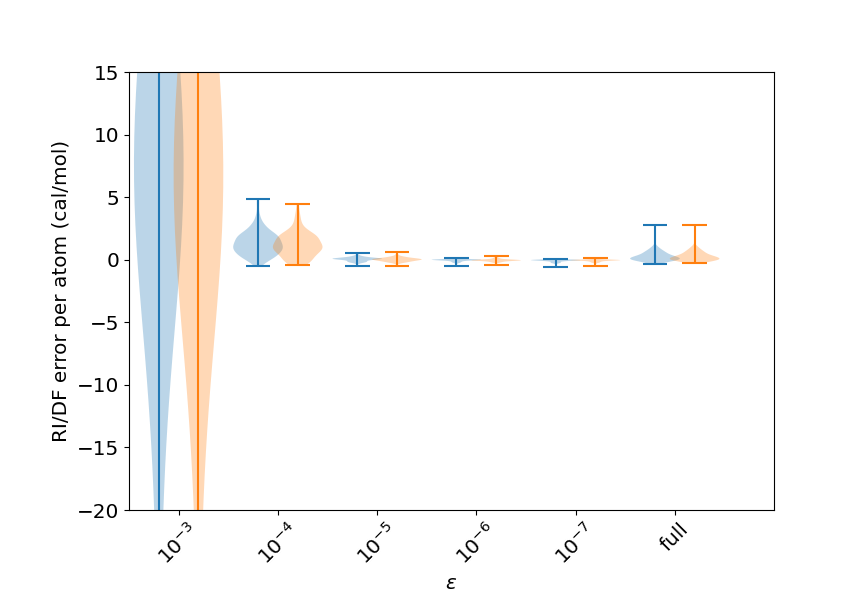}
\par\end{centering}
}

\caption{The effect of contraction on the RI/DF errors in the HF (cyan) and
MP2 (orange) total and atomization energies of the studied molecules
in the 3ZaPa-NR basis set. The last entry in the violin plot shows
the error distribution for the full, unpruned and uncontracted parent
auxiliary basis set. \label{fig:contr}}
\end{figure}

The RI/DF errors of the studied test set are shown as a function of
the contraction threshold in \ref{fig:contr} for HF and MP2 total
and atomization energies. The RI/DF errors converges rapidly when
the contraction threshold $\epsilon$ is decreased. We observe that
already the contraction threshold $\epsilon=10^{-4}$ affords RI/DF
errors in the order of a few $\mu E_{h}$ per electron in total energies,
and a few cal/mol per atom in atomization energies, which are negligible
compared to the truncation error of the OBS and the inherent errors
in the employed levels of theory.

It is also important to note here that the HF and MP2 errors behave
very similarly as a function of $\epsilon$. This result is fully
expected for the contraction based on a SVD of the three-index integrals:
because all integral classes are treated on the same footing, the
error is not expected to depend on the examined level of theory.

Interestingly, the RI/DF errors afforded by small values of $\epsilon$
are smaller than those obtained with the full parent auxiliary basis
set. This difference can likely be attributed to the better conditioning
of the contracted auxiliary basis set: unlike the parent ABS, the
functions of the contracted ABS are orthonormal on each atom, leading
to improved numerical stability in the RI/DF method when the contracted
ABS is used in polyatomic calculations; this in turn reduces numerical
errors. In addition, the contracted ABS also contains fewer functions,
which likewise improves numerical stability.

\begin{figure*}
\begin{centering}
\subfloat[Total energies, pruning with \ref{eq:lmax}]{\begin{centering}
\includegraphics[width=0.5\linewidth]{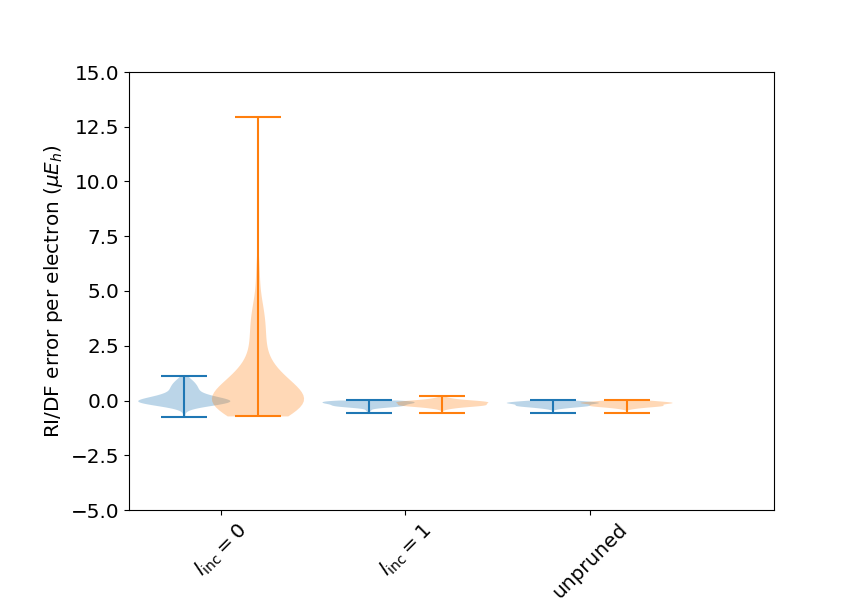}
\par\end{centering}
}\subfloat[Total energies, pruning with \ref{eq:lmax-yang}]{\begin{centering}
\includegraphics[width=0.5\linewidth]{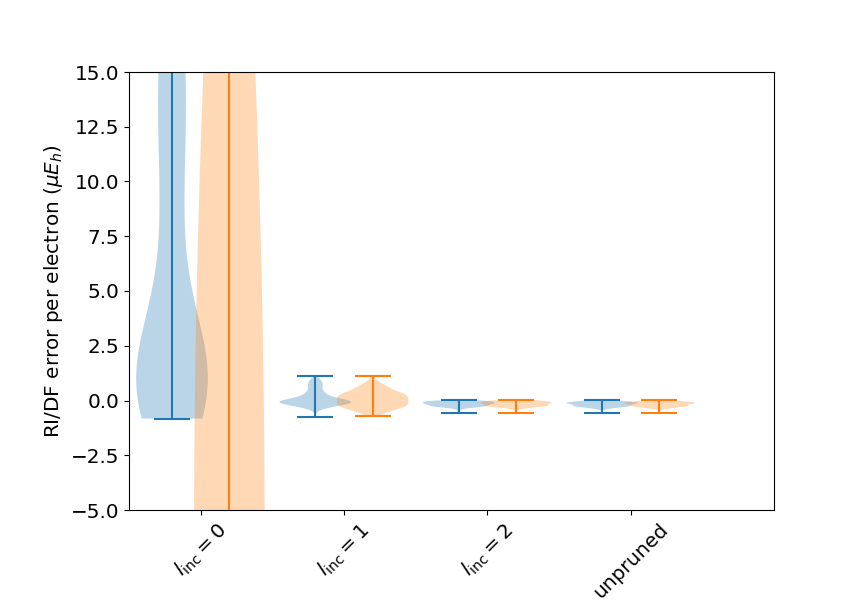}
\par\end{centering}
}
\par\end{centering}
\begin{centering}
\subfloat[Atomization energies, pruning with \ref{eq:lmax}]{\begin{centering}
\includegraphics[width=0.5\linewidth]{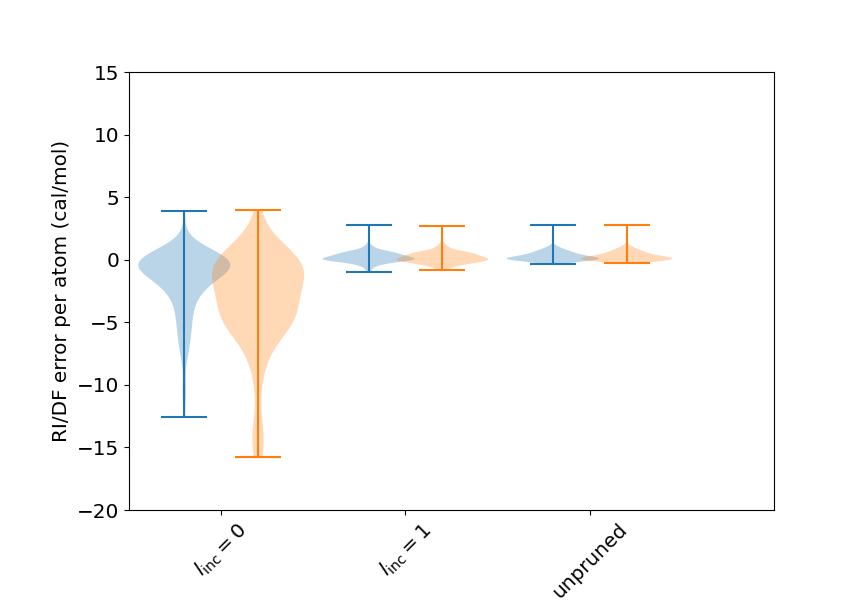}
\par\end{centering}
}\subfloat[Atomization energies, pruning with \ref{eq:lmax-yang}]{\begin{centering}
\includegraphics[width=0.5\linewidth]{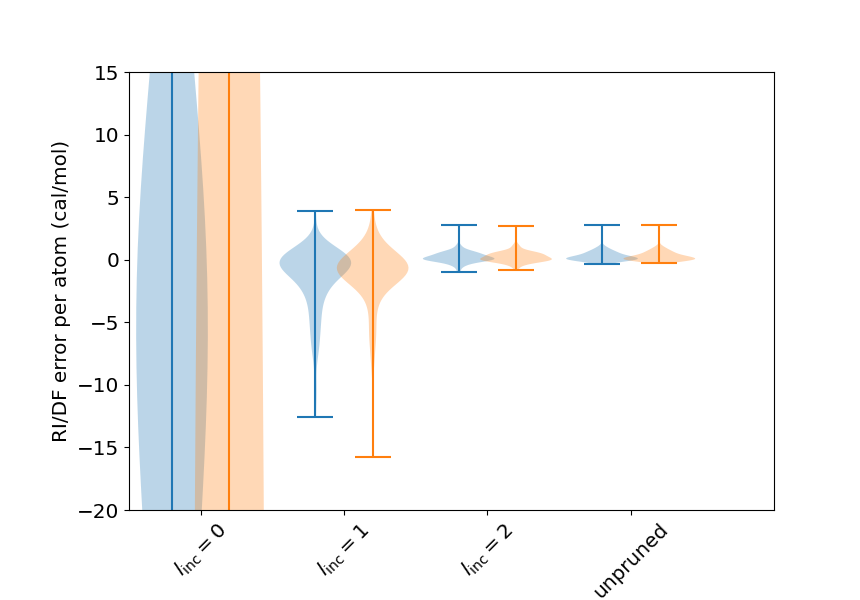}
\par\end{centering}
}
\par\end{centering}
\caption{The effect of pruning high-angular-momentum functions according to
\cref{eq:lmax,eq:lmax-yang} on the RI/DF errors HF (cyan) and MP2
(orange) total and atomization energies of the studied molecules in
the 3ZaPa-NR basis set. The last entries in the violin plots show
the error distribution for the full, unpruned and uncontracted parent
auxiliary basis set. \label{fig:trunc}}
\end{figure*}

We move on to studying the effect of pruning HAM auxiliary functions
according to \cref{eq:lmax,eq:lmax-yang}. The results of these procedures
on HF and MP2 total and atomization energies are shown in \ref{fig:trunc}. 

The value $l_{\text{inc}}=0$ yields good results with \ref{eq:lmax}:
although the MP2 total energy exhibits a small maximum absolute error
of some 13 $\mu E_{h}$ per electron (which arises for \ce{H2}),
the errors in HF total energies appear to be small and close to those
of the full original ABS. 

In contrast, the value $l_{\text{inc}}=0$ yields unacceptably large
errors for \ref{eq:lmax-yang}: for instance, the MP2 total energy
of the nitrogen atom is off by $-223\mu E_{h}$ per electron and the
MP2 atomization energy of \ce{P2} is off by $-893$ cal/mol per atom.

Because \ref{eq:lmax-yang} clearly exhibits the wrong physical behavior,
for the rest of this work we will consider only \ref{eq:lmax} for
pruning the HAM functions. With this scheme, $l_{\text{inc}}=0$ already
yields good accuracy, and the values with $l_{\text{inc}}=1$ are
almost converged to the full uncontracted and unpruned ABS, which
is reproduced by $l_{\text{inc}}=2$ for the presently considered
triple-$\zeta$ 3ZaPa-NR basis.

\subsection{Performance in Larger Basis Sets \label{subsec:Performance-in-larger}}

We observe from the analysis of \ref{subsec:3ZaPa-NR} that contraction
thresholds $\epsilon\lesssim10^{-4}$ afford suitably accurate total
and atomization energies, while pruning HAM functions with \ref{eq:lmax}
also works well. In this section, we examine the accuracy of the compound
procedure, where the ABS is both contracted and pruned of HAM functions. 

\begin{figure*}
\subfloat[3ZaPa-NR, $l_{\text{inc}}=0$]{\begin{centering}
\includegraphics[width=0.5\linewidth]{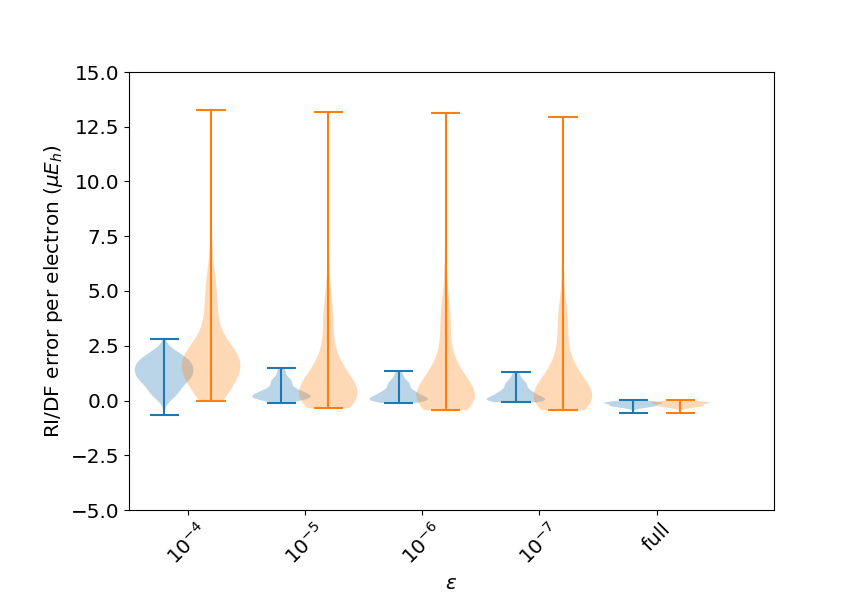}
\par\end{centering}
}\subfloat[3ZaPa-NR, $l_{\text{inc}}=1$]{\begin{centering}
\includegraphics[width=0.5\linewidth]{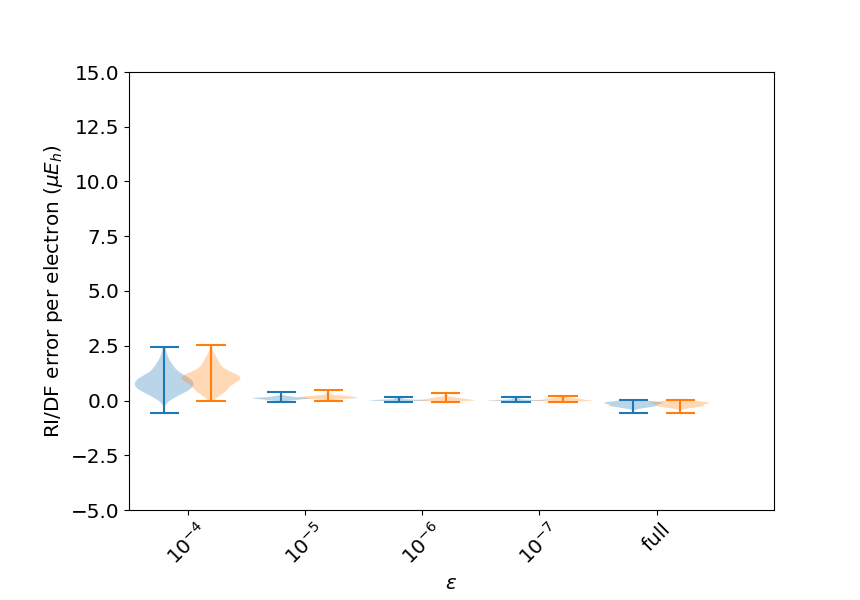}
\par\end{centering}
}

\subfloat[4ZaPa-NR, $l_{\text{inc}}=0$]{\begin{centering}
\includegraphics[width=0.5\linewidth]{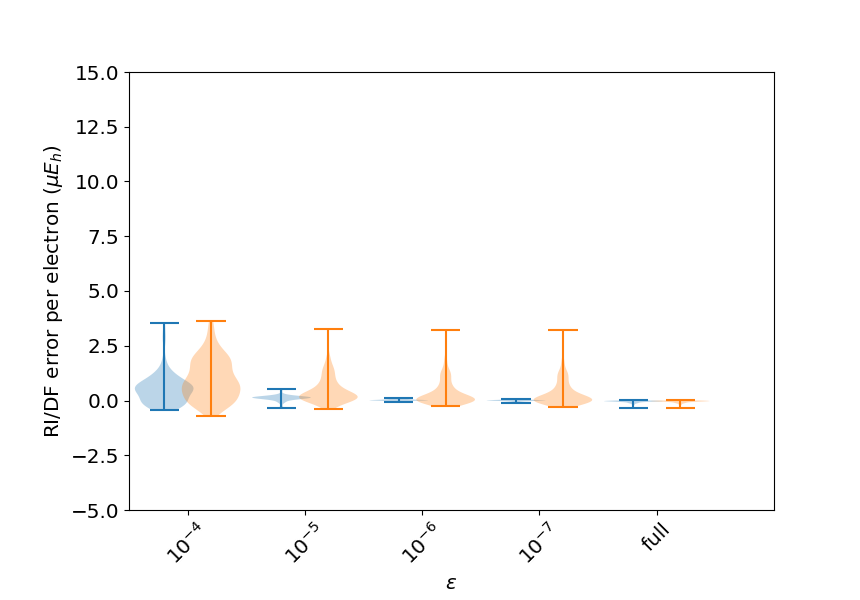}
\par\end{centering}
}\subfloat[4ZaPa-NR, $l_{\text{inc}}=1$]{\begin{centering}
\includegraphics[width=0.5\linewidth]{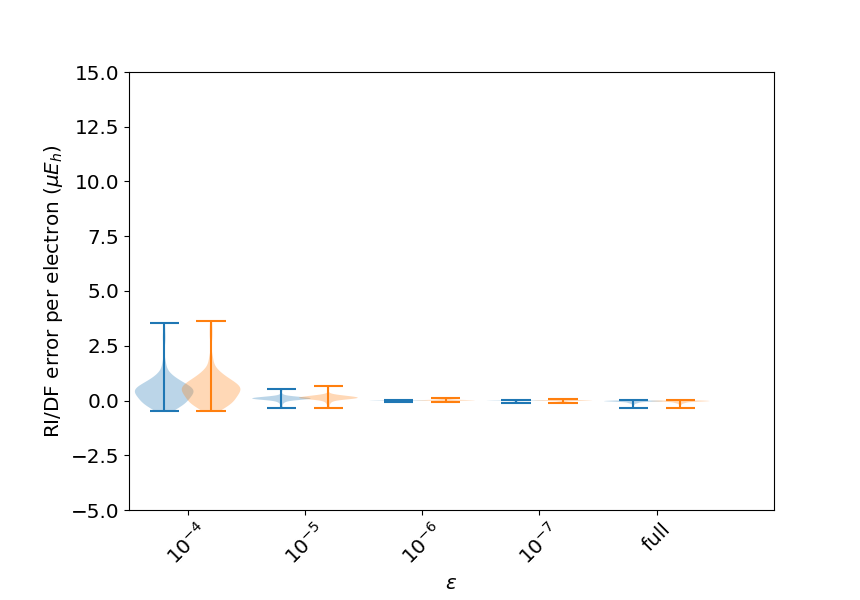}
\par\end{centering}
}

\subfloat[5ZaPa-NR, $l_{\text{inc}}=0$]{\begin{centering}
\includegraphics[width=0.5\linewidth]{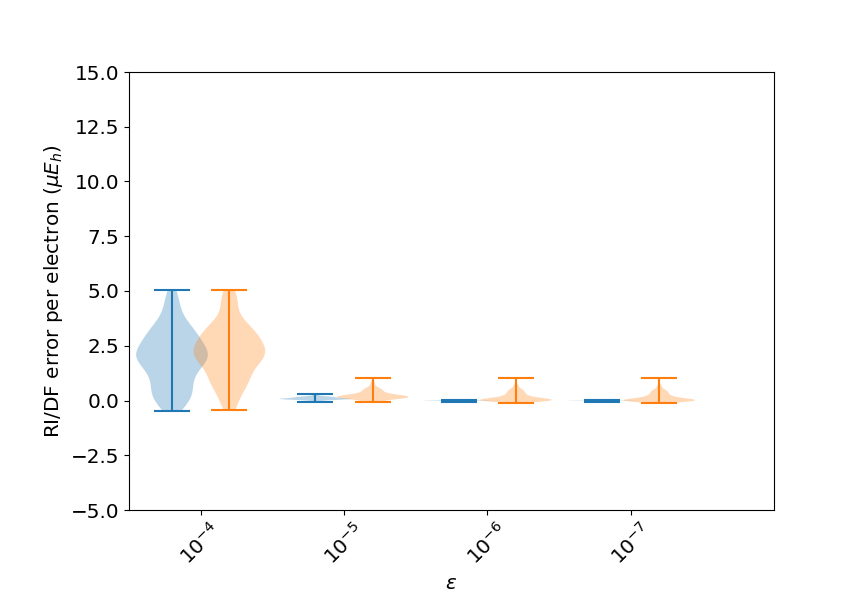}
\par\end{centering}
}\subfloat[5ZaPa-NR, $l_{\text{inc}}=1$]{\begin{centering}
\includegraphics[width=0.5\linewidth]{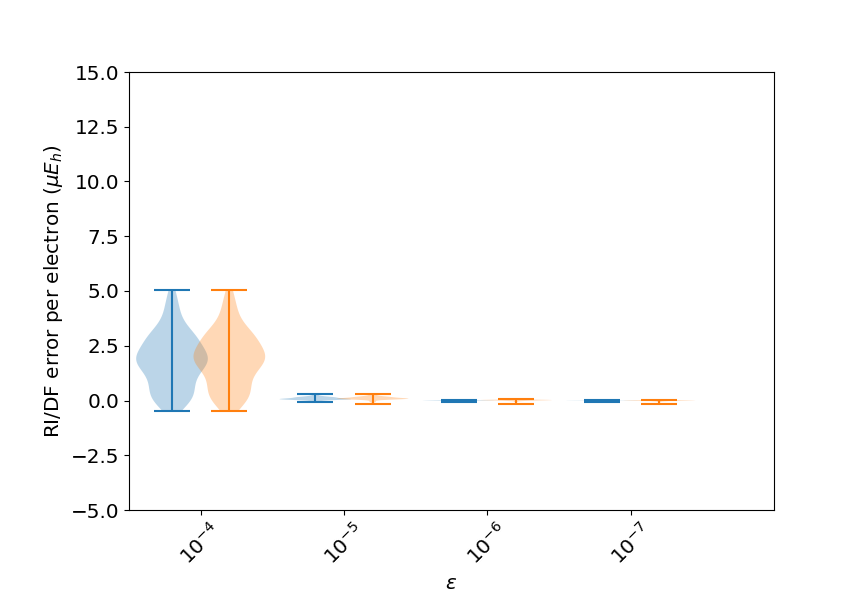}
\par\end{centering}
}

\caption{The effect of contraction on the RI/DF errors in the HF (cyan) and
MP2 (orange) total energies of the studied molecules in various basis
sets, when the auxiliary basis is also pruned with \ref{eq:lmax}
with $l_{\text{inc}}=0$ or $l_{\text{inc}}=1$. The last entry in
the 3ZaPa-NR and 4ZaPa-NR plots show the error distributions for the
full, unpruned and uncontracted parent auxiliary basis sets. \label{fig:newtrunc}}
\end{figure*}

Starting with the total energies shown in \ref{fig:newtrunc}, we
observe for all OBSs that the ABS contracted and pruned with \ref{eq:lmax}
with $l_{\text{inc}}=1$ affords negligible errors: as long as the
contraction threshold $\epsilon\lesssim10^{-5}$, the errors in the
HF and MP2 total energies are of the order of $1\ \mu E_{h}$ per
electron or smaller. 

We also observe that for a fixed value of $l_{\text{inc}}$, the maximum
RI/DF errors go down when the size of the OBS is increased. While
$l_{\text{inc}}=0$ leads to errors in the total energy of the order
of 13 $\mu E_{h}$ per electron with the triple-$\zeta$ 3ZaPa-NR
OBS, the error decreases to $\mathcal{O}(4\ \mu E_{h})$ per electron
in the quadruple-$\zeta$ 4ZaPa-NR OBS, and $\mathcal{O}(1\ \mu E_{h})$
per electron in the quintuple-$\zeta$ 5ZaPa-NR OBS. 

Given that the increase in the size of the OBS is accompanied with
an increase in the maximum angular momentum of the corresponding ABS
(\ref{eq:lmax}), a likely explanation for the decreasing error with
increasing OBS is that the one-center approximation that is the cornerstone
of the DF/RI method\citep{Pedersen2023__} is becoming more accurate:
energetically important two-center products of OBS functions are captured
to better and better accuracy, when the ABS becomes larger and larger.
Increasing the cardinal number of the OBS adds higher polarization
shells in the autogenerated ABS, but it also adds more radial functions
to lower angular momenta in the ABS.

\begin{figure*}
\subfloat[3ZaPa-NR, $l_{\text{inc}}=0$]{\begin{centering}
\includegraphics[width=0.5\linewidth]{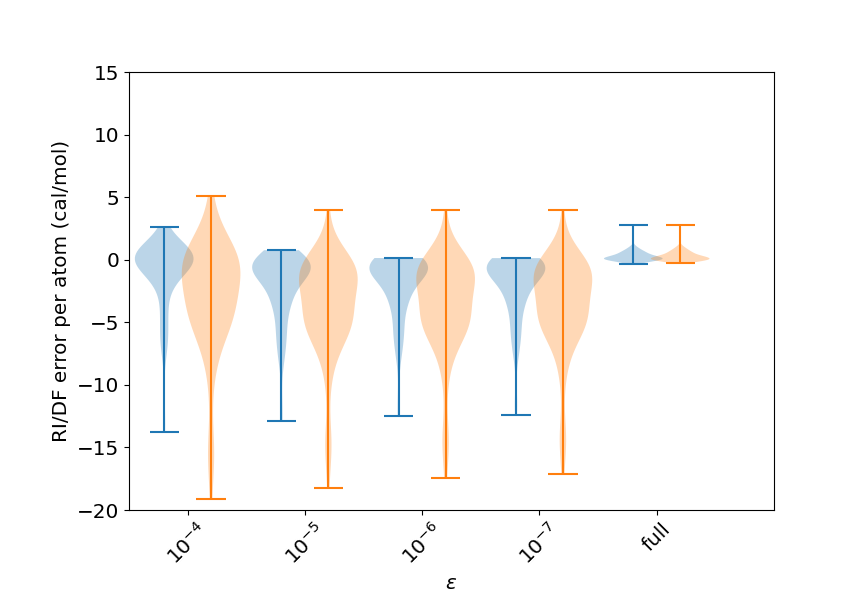}
\par\end{centering}
}\subfloat[3ZaPa-NR, $l_{\text{inc}}=1$]{\begin{centering}
\includegraphics[width=0.5\linewidth]{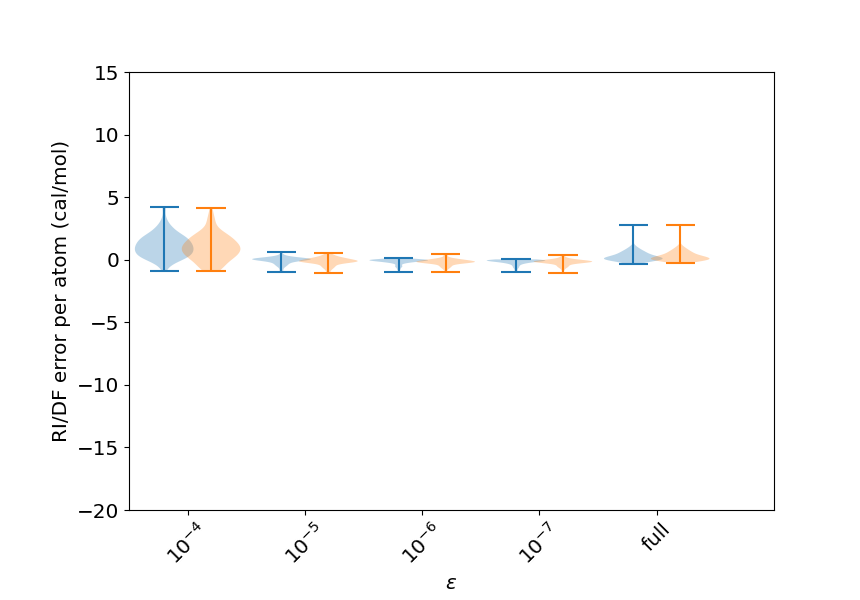}
\par\end{centering}
}

\subfloat[4ZaPa-NR, $l_{\text{inc}}=0$]{\begin{centering}
\includegraphics[width=0.5\linewidth]{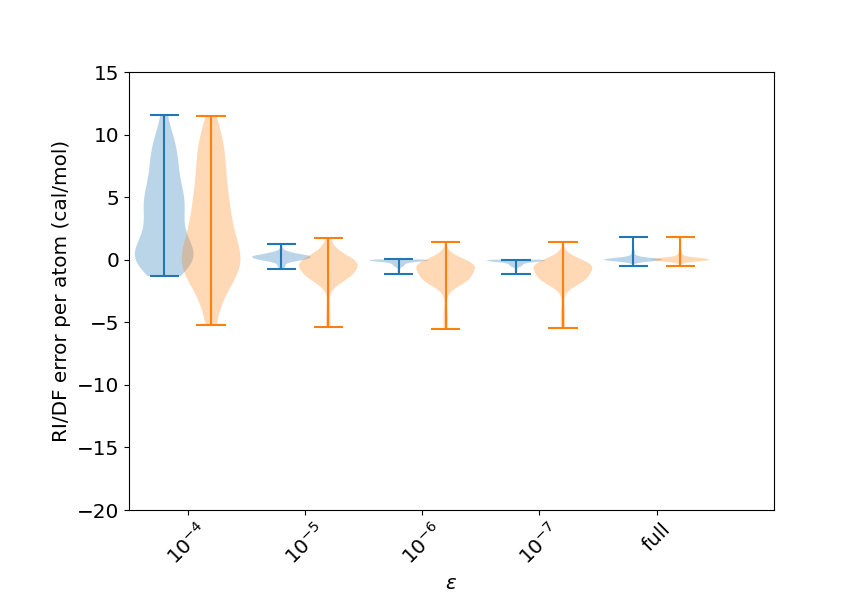}
\par\end{centering}
}\subfloat[4ZaPa-NR, $l_{\text{inc}}=1$]{\begin{centering}
\includegraphics[width=0.5\linewidth]{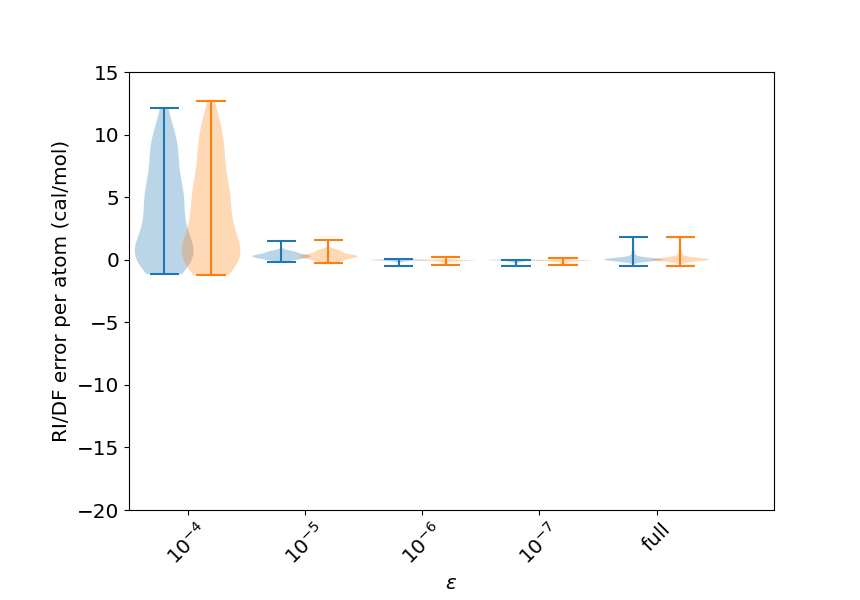}
\par\end{centering}
}

\subfloat[5ZaPa-NR, $l_{\text{inc}}=0$]{\begin{centering}
\includegraphics[width=0.5\linewidth]{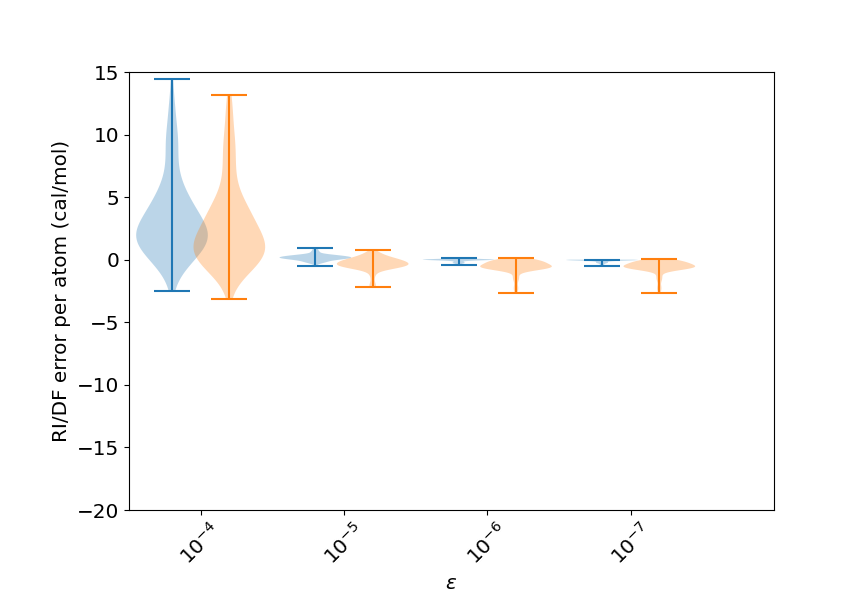}
\par\end{centering}
}\subfloat[5ZaPa-NR, $l_{\text{inc}}=1$]{\begin{centering}
\includegraphics[width=0.5\linewidth]{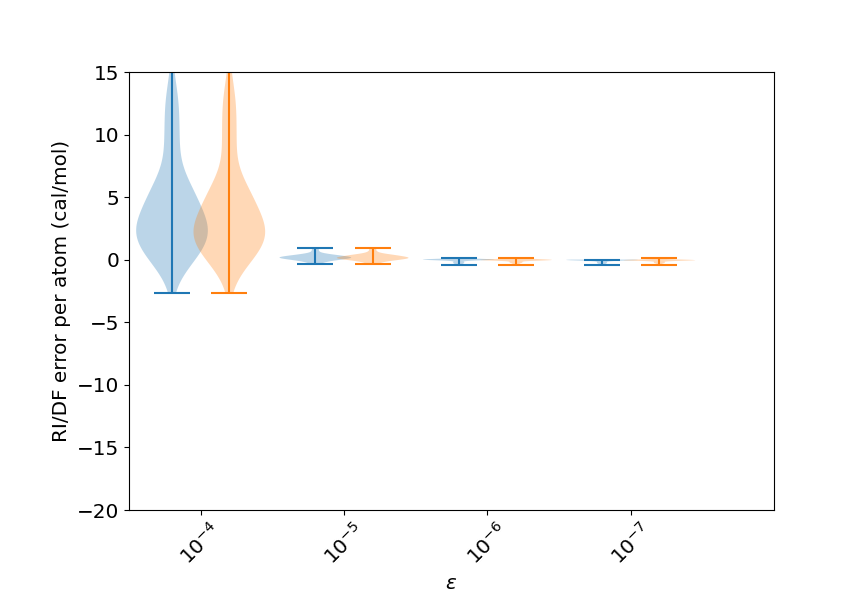}
\par\end{centering}
}

\caption{The effect of contraction on the RI/DF errors in the HF (cyan) and
MP2 (orange) atomization energies of the studied molecules in various
basis sets, when the auxiliary basis is also pruned with \ref{eq:lmax}
with $l_{\text{inc}}=0$ or $l_{\text{inc}}=1$. The last entry in
the 3ZaPa-NR and 4ZaPa-NR plots show the error distributions for the
full, unpruned and uncontracted parent auxiliary basis sets. \label{fig:newtrunc-atomization}}
\end{figure*}

The corresponding results for atomization energies are shown in \ref{fig:newtrunc-atomization}.
The errors in atomization energies of the order of cal/mol per atom
are orders of magnitude smaller than the usual limit of chemical accuracy
of 1 \textbf{k}cal/mol. We again note the increasing accuracy of the
ABS with fixed $l_{\text{inc}}$ in increasing size of the OBS, which
is clear in the small-$\epsilon$ data for $l_{\text{inc}}=0$. Excellent
accuracy is achieved with $l_{\text{inc}}=1$ and $\epsilon\lesssim10^{-5}$
in all OBSs. 

\subsection{Size Reductions \label{subsec:Size-reductions}}

Having established that an excellent level of accuracy can be reached
with the DF/RI approximation with the presently considered pruning
and contraction technique for the automatically generated ABSs, we
can proceed to discussing the associated reductions in the size of
the ABS. As was already mentioned above, the computational cost of
the DF/RI technique is linear in the size of the ABS; therefore, discussion
on the cost of the DF/RI technique usually boils down to the examination
of the ratio 
\begin{equation}
\gamma=N_{\text{ABS}}^{\text{bf}}/N_{\text{OBS}}^{\text{bf}},\label{eq:gamma}
\end{equation}
that is, the number of auxiliary functions divided by the number of
orbital functions. 

In the following, we will examine the accuracy and cost of automatically
generated ABSs with the present method, considering simultaneous contraction
with $\epsilon\in[10^{-4},10^{-5},10^{-6}]$ and pruning with $l_{\text{inc}}\in[0,1]$,
which yielded good results above in \ref{subsec:Performance-in-larger}.
We will compare the results to those of the AutoAux procedure of \citet{Stoychev2017_JCTC_554}. 

\begin{figure*}
\subfloat[3ZaPa-NR total energies]{\begin{centering}
\includegraphics[width=0.5\linewidth]{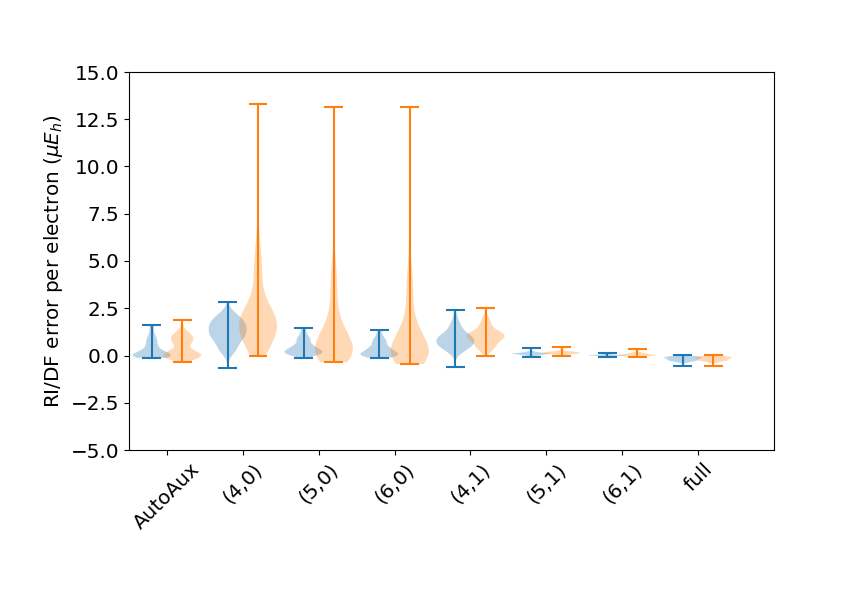}
\par\end{centering}
}\subfloat[3ZaPa-NR atomization energies]{\begin{centering}
\includegraphics[width=0.5\linewidth]{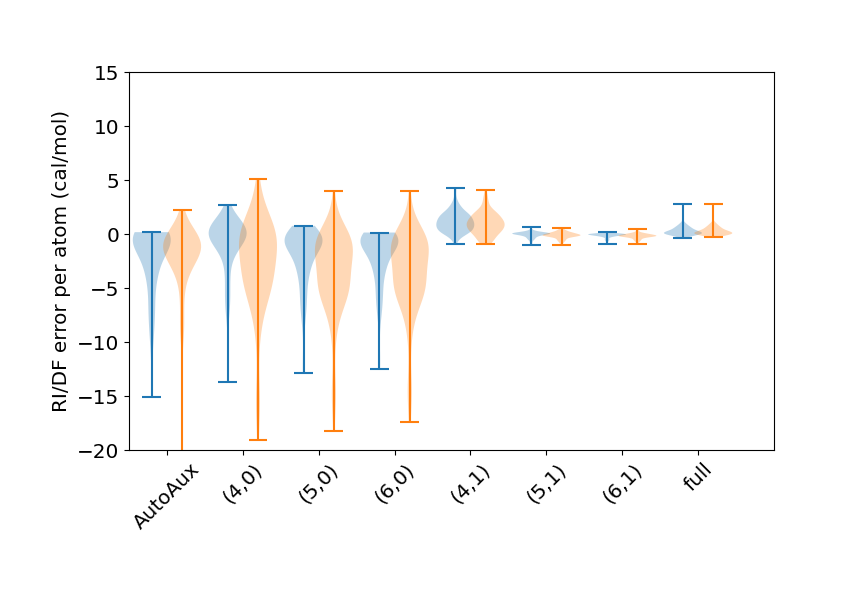}
\par\end{centering}
}

\subfloat[4ZaPa-NR total energies]{\begin{centering}
\includegraphics[width=0.5\linewidth]{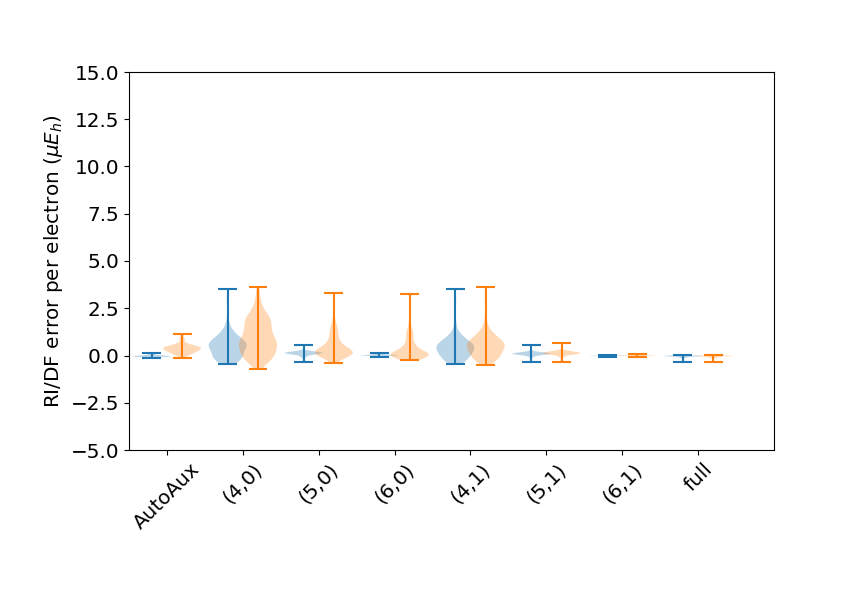}
\par\end{centering}
}\subfloat[4ZaPa-NR atomization energies]{\begin{centering}
\includegraphics[width=0.5\linewidth]{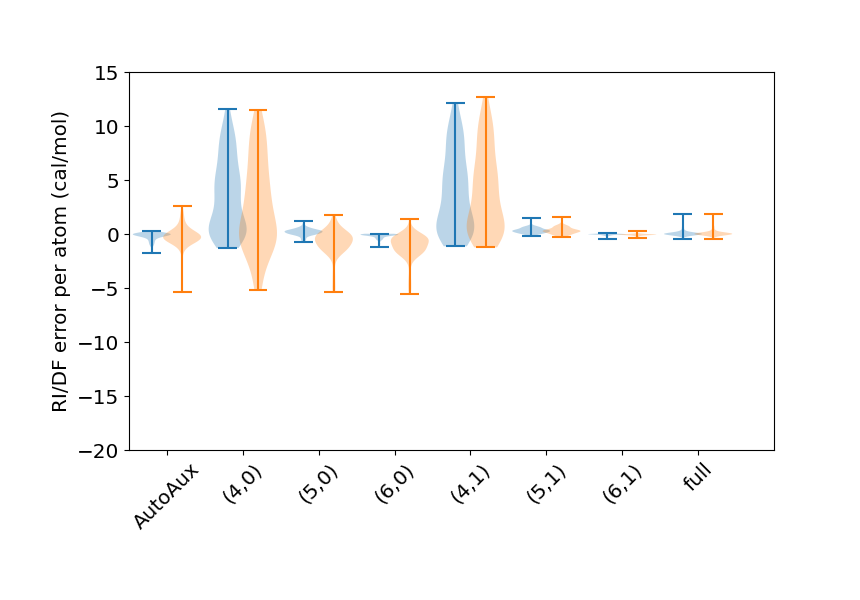}
\par\end{centering}
}

\subfloat[5ZaPa-NR total energies]{\begin{centering}
\includegraphics[width=0.5\linewidth]{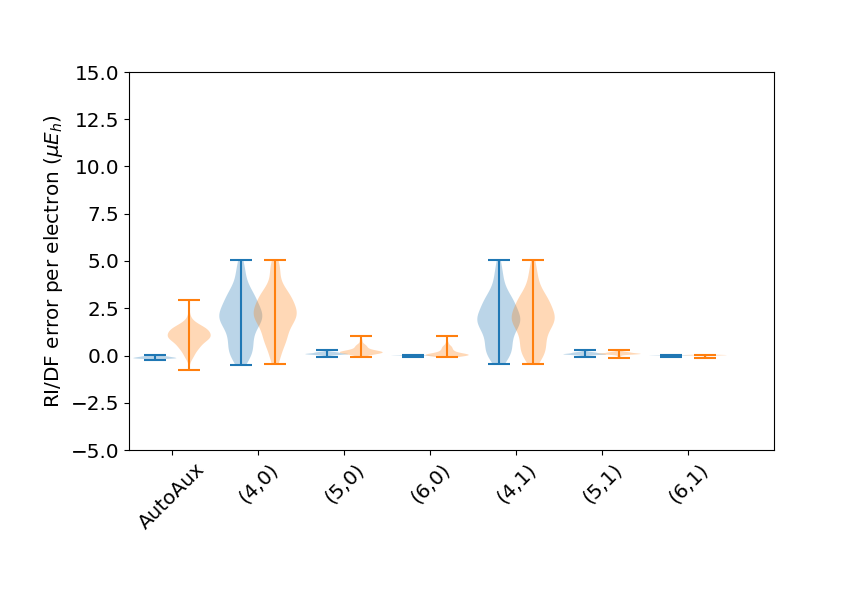}
\par\end{centering}
}\subfloat[5ZaPa-NR atomization energies]{\begin{centering}
\includegraphics[width=0.5\linewidth]{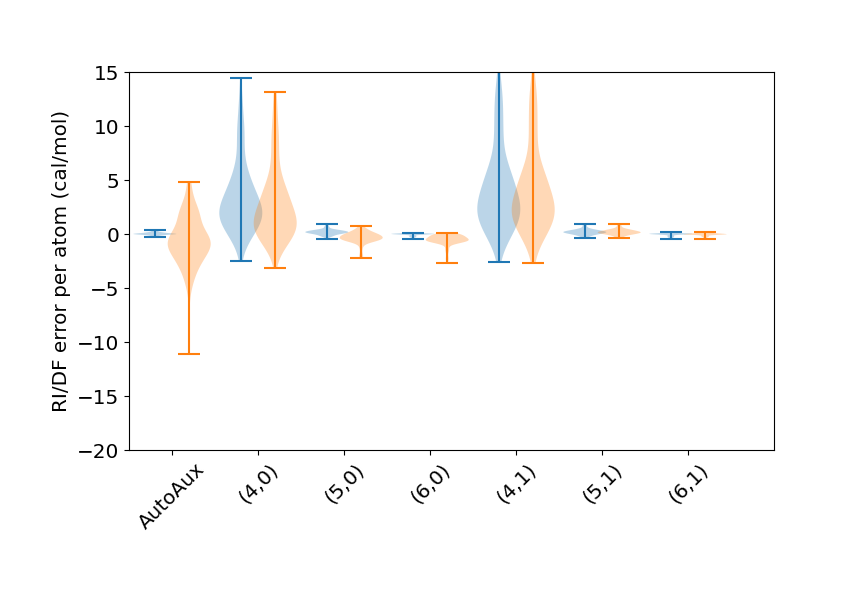}
\par\end{centering}
}

\caption{Summary of the performance of the various automated schemes for automated
auxiliary basis set construction across various orbital basis sets:
the AutoAux procedure of \citet{Stoychev2017_JCTC_554}, as well as
contracted and pruned basis sets with the parameters $\epsilon=10^{-x}$
and $l_{\text{inc}}=y$ specified with the shorthand $(x,y)$. The
last entry in the 3ZaPa-NR and 4ZaPa-NR plots show the error distributions
for the full, unpruned and uncontracted parent auxiliary basis sets.
\label{fig:Performance-of-the}}
\end{figure*}

The accuracy of these ABSs is shown in \ref{fig:Performance-of-the}.
From these data, we can identify reasonable combinations of the truncation
parameters, which appear to be balanced in the contraction and pruning
of HAM functions. The choice $\epsilon=10^{-4}$ and $l_{\text{inc}}=0$
yields a ``small'' ABS, while $\epsilon=10^{-5}$ and $l_{\text{inc}}=1$
leads to a ``large'' ABS, and $\epsilon=10^{-6}$ and $l_{\text{inc}}=1$
yields a ``verylarge'' ABS. 

The AutoAux basis is much more accurate for HF than for MP2 in large
OBSs, and the method does not offer a straightforward way to improve
the accuracy of the generated ABS. In contrast, the ``small'', ``large''
and ``verylarge'' ABSs obtained with the above procedure provide
a sequence of improving accuracy in all studied OBSs. With the exclusion
of the ``small'' ABS in the 3ZaPa-NR OBS, they also afford similar
accuracy at both HF and MP2 levels of theory. 

\begin{figure*}
\subfloat[3ZaPa-NR]{\begin{centering}
\includegraphics[width=1\linewidth]{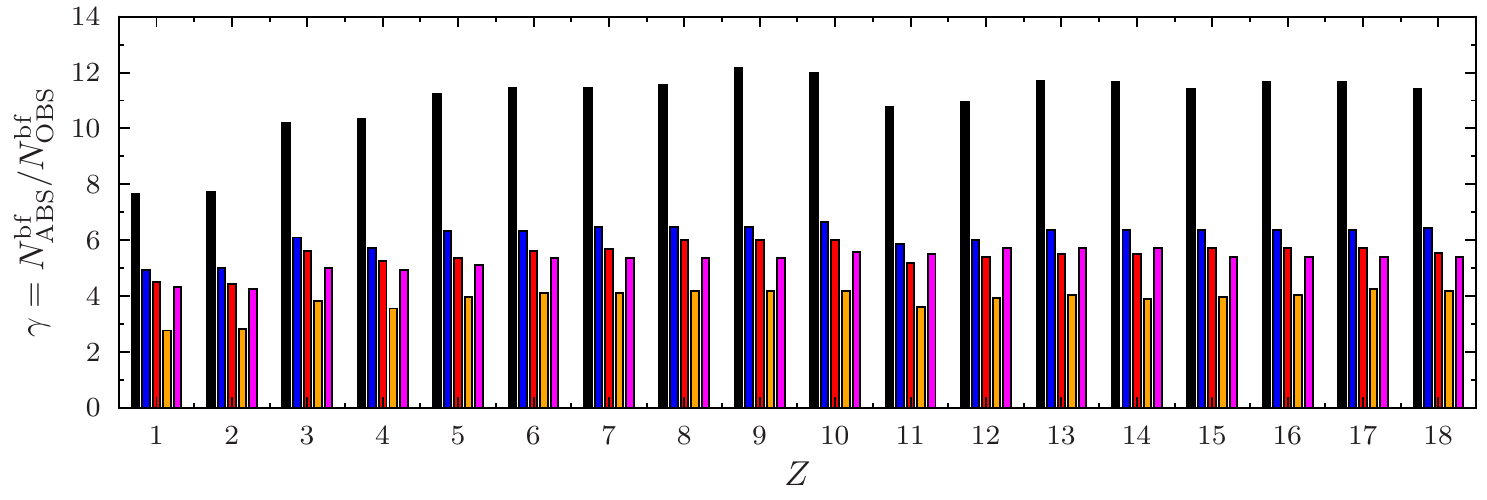}
\par\end{centering}
}

\subfloat[4ZaPa-NR]{\begin{centering}
\includegraphics[width=1\linewidth]{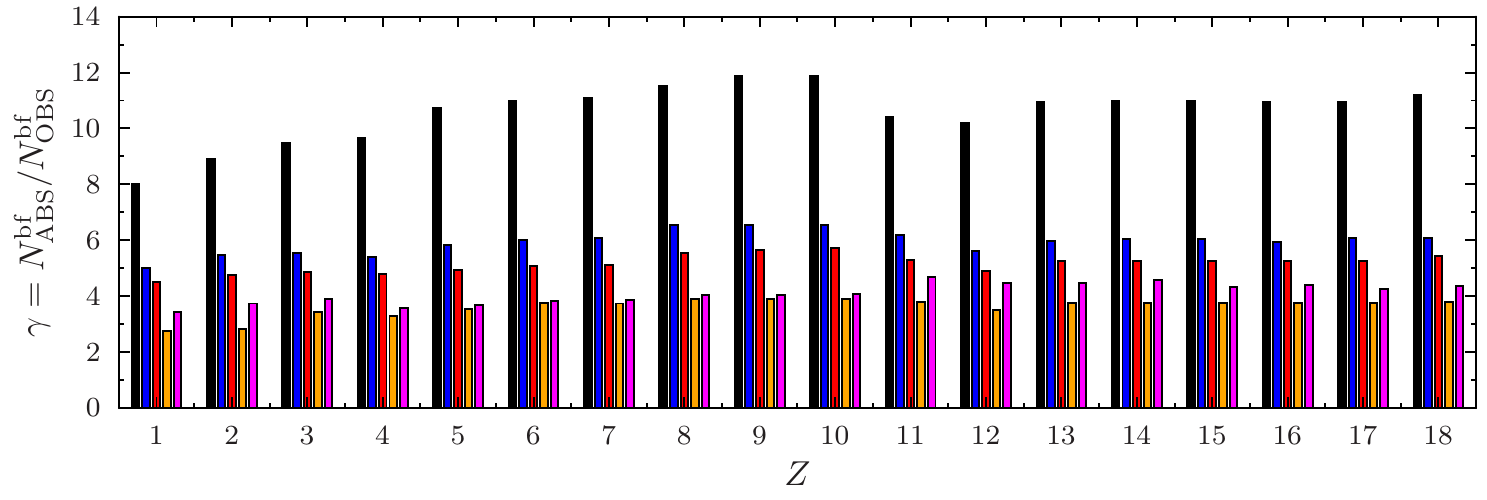}
\par\end{centering}
}

\subfloat{\centering{}\includegraphics[width=1\linewidth]{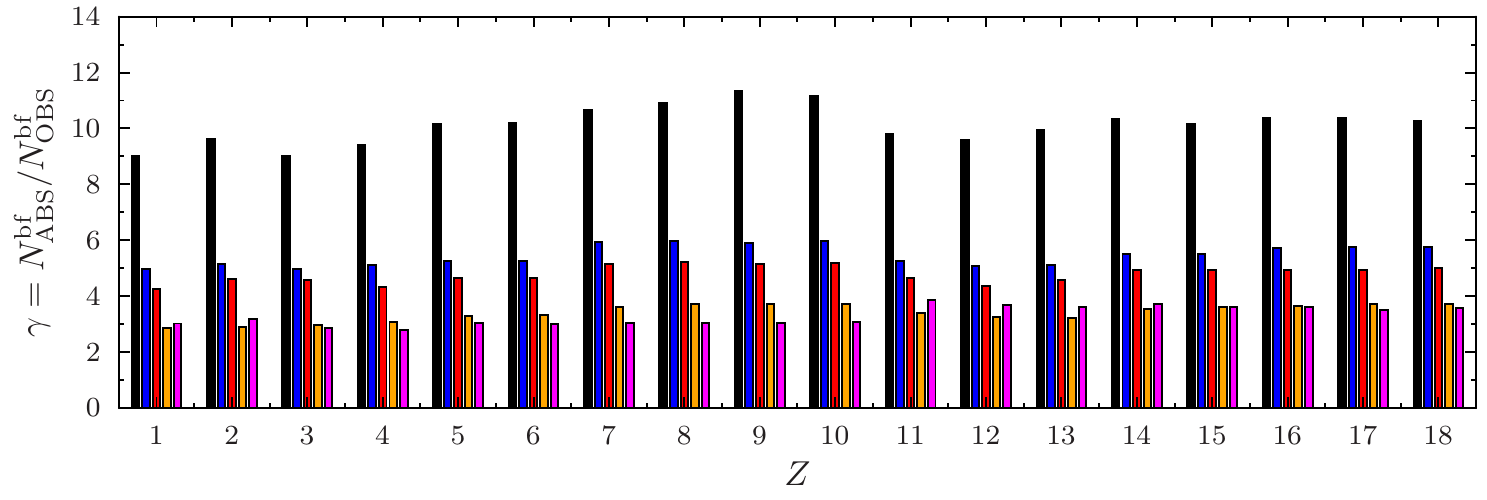}}

\caption{Sizes of autogenerated auxiliary basis for 3ZaPa-NR, 4ZaPa-NR, and
5ZaPa-NR for H--Ar. Reading in order from the left, the original
primitive auxiliary basis set, generated by the procedure of \citeref{Lehtola2021_JCTC_6886}
is shown in black. The verylarge ($\epsilon=10^{-6}$ and $l_{\text{inc}}=1$),
large ($\epsilon=10^{-5}$ and $l_{\text{inc}}=1$), and small ($\epsilon=10^{-4}$
and $l_{\text{inc}}=0$ ) auxiliary sets are shown in blue, red, and
orange, respectively. For comparison, the size of the ABS generated
with the AutoAux procedure of \citet{Stoychev2017_JCTC_554} is shown
in magenta. \label{fig:Size-of-autogenerated}}
\end{figure*}

The ratios $\gamma$ defined by \ref{eq:gamma} for the various ABSs
are shown in \ref{fig:Size-of-autogenerated}, where the size of the
full, unpruned and uncontracted autogenerated ABS is also shown for
reference. The minimum and maximum values of $\gamma$ are also given
in \ref{tab:Size-range-of} for all OBSs and the corresponding ABSs. 

As was already discussed in \citeref{Lehtola2021_JCTC_6886}, the
full autogenerated auxiliary basis sets are large: we see from \ref{fig:Size-of-autogenerated}
and \ref{tab:Size-range-of} that the ratio $\gamma$ varies from
8 to 12 in the full ABS. However, we also see that the contraction
and pruning procedure of this work leads to significant reductions
in the size of the ABSs. 

Already going from the full ABS to the ``verylarge'' ABS is seen
to roughly halve the number of auxiliary functions, indicating potential
for significant savings with negligible errors as shown by the related
accuracy data in \ref{fig:Performance-of-the}. 

A further reduction is achieved by going to the ``large'' ABS, which
still affords consistently small errors (\ref{fig:Performance-of-the}).
The ``small'' ABS, in turn, exhibits $\gamma$ in the range 3--4
for all OBSs, enabling quick calculations with somewhat larger errors,
which may still be sufficiently small for practical applications.

\begin{table*}
\begin{centering}
\begin{tabular}{c|cccccccc|cc}
 & \multicolumn{2}{c}{full} & \multicolumn{2}{c}{``verylarge''} & \multicolumn{2}{c}{``large''} & \multicolumn{2}{c|}{``small''} & \multicolumn{2}{c}{AutoAux}\tabularnewline
 & $\min\gamma$ & $\max\gamma$ & $\min\gamma$ & $\max\gamma$ & $\min\gamma$ & $\max\gamma$ & $\min\gamma$ & $\max\gamma$ & $\min\gamma$ & $\max\gamma$\tabularnewline
\hline 
\hline 
3ZaPa-NR & 7.7 & 12.2 & 4.9 & 6.7 & 4.4 & 6.0 & 2.8 & 4.2 & 4.3 & 5.7\tabularnewline
4ZaPa-NR & 8.0 & 11.9 & 5.0 & 6.5 & 4.5 & 5.7 & 2.8 & 3.9 & 3.5 & 4.7\tabularnewline
5ZaPa-NR & 9.0 & 11.3 & 5.0 & 6.0 & 4.3 & 5.2 & 2.9 & 3.7 & 2.8 & 3.9\tabularnewline
\end{tabular}
\par\end{centering}
\caption{Ranges of $\gamma$ defined by \ref{eq:gamma} of various automatically
generated ABSs.\label{tab:Size-range-of}}

\end{table*}

\subsection{Role of Primitive Auxiliary Basis}

Having demonstrated the accuracy of the reduction scheme, we examine
the role of the primitive auxiliary basis fed into the contraction
procedure. We will compare contracted ABSs formed with and without
the prescreening of the orbital products, which was performed for
the results above with a pivoted Cholesky decomposition of the ERI
tensor as recommended in \citeref{Lehtola2021_JCTC_6886}.

Comparing the compositions of the contracted ABSs for the 3ZaPa-NR
and 4ZaPa-NR OBSs shown in \ref{tab:Compositions-of-contracted},
we observe that despite the large differences in the number of Gaussian
primitives in the ABSs generated with and without the orbital product
screening---which were already observed in \citeref{Lehtola2021_JCTC_6886}---in
both cases the contracted ABSs turn out to have the same composition,
that is, the same number of contracted functions.

However, the use of the prescreening technique is still highly attractive
for GTO calculations. Because the integrals are evaluated in terms
of the primitive GTOs, reducing the number of primitive functions
is key to making the approach computationally efficient.

But, when the present contraction scheme is used with NAOs, it appears
that prescreening the orbital products via a pivoted Cholesky decomposition
of the ERI tensor is indeed unnecessary: the simpler scheme of simply
forming all radial orbital products and recontracting them leads to
the same compact ABS at the same computational cost. The key difference
to GTOs is that NAO integrals are evaluated by quadrature, and the
contraction can be carried out as a transformation of the NAOs' expansion
coefficients; see \citeref{Lehtola2019_IJQC_25945} for details.

\begin{table*}
\subfloat[3ZaPa-NR]{\begin{centering}
\begin{tabular}{ccccc}
 & \multicolumn{2}{c}{{\footnotesize{}without prescreening}} & \multicolumn{2}{c}{{\footnotesize{}with prescreening}}\tabularnewline
 & {\footnotesize{}primitive} & {\footnotesize{}contracted} & {\footnotesize{}primitive} & {\footnotesize{}contracted}\tabularnewline
\hline 
\hline 
{\footnotesize{}H} & {\footnotesize{}15s14p10d4f1g} & {\footnotesize{}9s7p6d3f1g} & {\footnotesize{}15s11p10d4f1g} & {\footnotesize{}9s7p6d3f1g}\tabularnewline
{\footnotesize{}He} & {\footnotesize{}17s16p11d4f1g} & {\footnotesize{}8s7p6d3f1g} & {\footnotesize{}16s12p9d4f1g} & {\footnotesize{}8s7p6d3f1g}\tabularnewline
{\footnotesize{}Li} & {\footnotesize{}24s23p20d17f9g4h1i} & {\footnotesize{}11s9p9d7f6g3h1i} & {\footnotesize{}22s19p16d12f9g4h1i} & {\footnotesize{}11s9p9d7f6g3h1i}\tabularnewline
{\footnotesize{}Be} & {\footnotesize{}24s23p20d18f9g4h1i} & {\footnotesize{}11s9p8d7f5g3h1i} & {\footnotesize{}24s20p15d13f9g4h1i} & {\footnotesize{}11s9p8d7f5g3h1i}\tabularnewline
{\footnotesize{}B} & {\footnotesize{}24s25p22d20f10g4h1i} & {\footnotesize{}10s9p9d7f5g3h1i} & {\footnotesize{}23s21p18d15f10g4h1i} & {\footnotesize{}10s9p9d7f5g3h1i}\tabularnewline
{\footnotesize{}C} & {\footnotesize{}26s25p23d20f10g4h1i} & {\footnotesize{}11s9p9d7f6g3h1i} & {\footnotesize{}23s21p18d14f10g4h1i} & {\footnotesize{}11s9p9d7f6g3h1i}\tabularnewline
{\footnotesize{}N} & {\footnotesize{}25s25p23d21f10g4h1i} & {\footnotesize{}11s10p9d7f6g3h1i} & {\footnotesize{}24s23p20d14f9g4h1i} & {\footnotesize{}11s10p9d7f6g3h1i}\tabularnewline
{\footnotesize{}O} & {\footnotesize{}25s26p23d21f11g4h1i} & {\footnotesize{}12s10p10d8f6g3h1i} & {\footnotesize{}25s23p20d15f9g4h1i} & {\footnotesize{}12s10p10d8f6g3h1i}\tabularnewline
{\footnotesize{}F} & {\footnotesize{}26s26p23d23f11g4h1i} & {\footnotesize{}12s10p10d8f6g3h1i} & {\footnotesize{}25s23p21d16f10g4h1i} & {\footnotesize{}12s10p10d8f6g3h1i}\tabularnewline
{\footnotesize{}Ne} & {\footnotesize{}26s26p24d22f11g4h1i} & {\footnotesize{}12s10p10d8f6g3h1i} & {\footnotesize{}25s24p22d16f10g4h1i} & {\footnotesize{}12s10p10d8f6g3h1i}\tabularnewline
{\footnotesize{}Na} & {\footnotesize{}31s29p28d26f15g5h1i} & {\footnotesize{}13s10p10d7f7g4h1i} & {\footnotesize{}29s27p23d15f11g5h1i} & {\footnotesize{}13s10p10d7f7g4h1i}\tabularnewline
{\footnotesize{}Mg} & {\footnotesize{}30s32p29d28f15g5h1i} & {\footnotesize{}14s11p10d8f7g4h1i} & {\footnotesize{}30s28p24d16f11g5h1i} & {\footnotesize{}14s11p10d8f7g4h1i}\tabularnewline
{\footnotesize{}Al} & {\footnotesize{}30s32p31d29f16g5h1i} & {\footnotesize{}14s11p11d8f7g4h1i} & {\footnotesize{}30s30p26d18f11g5h1i} & {\footnotesize{}14s11p11d8f7g4h1i}\tabularnewline
{\footnotesize{}Si} & {\footnotesize{}31s31p29d29f16g5h1i} & {\footnotesize{}14s11p11d8f7g4h1i} & {\footnotesize{}30s29p25d18f12g5h1i} & {\footnotesize{}14s11p11d8f7g4h1i}\tabularnewline
{\footnotesize{}P} & {\footnotesize{}31s31p30d28f16g5h1i} & {\footnotesize{}14s12p11d9f7g4h1i} & {\footnotesize{}30s29p25d18f12g5h1i} & {\footnotesize{}14s12p11d9f7g4h1i}\tabularnewline
{\footnotesize{}S} & {\footnotesize{}30s31p29d28f16g5h1i} & {\footnotesize{}14s12p11d9f7g4h1i} & {\footnotesize{}30s30p26d18f12g5h1i} & {\footnotesize{}14s12p11d9f7g4h1i}\tabularnewline
{\footnotesize{}Cl} & {\footnotesize{}30s31p29d28f16g5h1i} & {\footnotesize{}14s12p11d9f7g4h1i} & {\footnotesize{}30s29p26d19f12g5h1i} & {\footnotesize{}14s12p11d9f7g4h1i}\tabularnewline
{\footnotesize{}Ar} & {\footnotesize{}30s31p29d28f15g5h1i} & {\footnotesize{}13s12p11d8f7g4h1i} & {\footnotesize{}29s30p25d18f12g5h1i} & {\footnotesize{}13s12p11d8f7g4h1i}\tabularnewline
\end{tabular}
\par\end{centering}
}

\subfloat[4ZaPa-NR]{\begin{centering}
\begin{tabular}{ccccc}
 & \multicolumn{2}{c}{{\footnotesize{}without prescreening}} & \multicolumn{2}{c}{{\footnotesize{}with prescreening}}\tabularnewline
 & {\footnotesize{}primitive} & {\footnotesize{}contracted} & {\footnotesize{}primitive} & {\footnotesize{}contracted}\tabularnewline
\hline 
\hline 
{\footnotesize{}H} & {\footnotesize{}18s16p15d12f8g4h1i} & {\footnotesize{}11s9p8d7f6g3h1i} & {\footnotesize{}16s13p13d11f7g4h1i} & {\footnotesize{}11s9p8d7f6g3h1i}\tabularnewline
{\footnotesize{}He} & {\footnotesize{}20s19p18d14f8g4h1i} & {\footnotesize{}10s9p9d7f6g3h1i} & {\footnotesize{}20s15p13d12f8g4h1i} & {\footnotesize{}11s9p9d7f6g3h1i}\tabularnewline
{\footnotesize{}Li} & {\footnotesize{}26s25p23d22f19g11h7i4j1k} & {\footnotesize{}13s11p10d8f7g6h5i3j1k} & {\footnotesize{}25s21p17d14f13g10h7i4j1k} & {\footnotesize{}13s11p10d8f7g6h5i3j1k}\tabularnewline
{\footnotesize{}Be} & {\footnotesize{}26s25p22d23f21g11h7i4j1k} & {\footnotesize{}11s10p10d8f7g6h5i3j1k} & {\footnotesize{}26s21p17d14f12g10h7i4j1k} & {\footnotesize{}11s10p10d8f7g6h5i3j1k}\tabularnewline
{\footnotesize{}B} & {\footnotesize{}27s27p24d24f22g12h8i4j1k} & {\footnotesize{}11s11p10d9f7g6h5i3j1k} & {\footnotesize{}26s23p19d17f15g11h8i4j1k} & {\footnotesize{}11s11p10d9f7g6h5i3j1k}\tabularnewline
{\footnotesize{}C} & {\footnotesize{}28s27p25d24f23g13h8i4j1k} & {\footnotesize{}11s11p10d9f8g6h5i3j1k} & {\footnotesize{}28s23p20d17f15g11h8i4j1k} & {\footnotesize{}11s11p10d9f8g6h5i3j1k}\tabularnewline
{\footnotesize{}N} & {\footnotesize{}29s28p27d25f25g14h8i4j1k} & {\footnotesize{}13s11p12d10f8g6h5i3j1k} & {\footnotesize{}28s26p23d18f15g12h8i4j1k} & {\footnotesize{}13s11p12d10f8g6h5i3j1k}\tabularnewline
{\footnotesize{}O} & {\footnotesize{}28s28p26d25f25g14h9i4j1k} & {\footnotesize{}13s12p11d9f8g7h6i3j1k} & {\footnotesize{}28s25p22d18f15g12h9i4j1k} & {\footnotesize{}13s12p11d9f8g7h6i3j1k}\tabularnewline
{\footnotesize{}F} & {\footnotesize{}28s29p27d26f25g15h9i4j1k} & {\footnotesize{}13s12p11d10f8g7h6i3j1k} & {\footnotesize{}27s27p23d19f16g12h8i4j1k} & {\footnotesize{}13s12p11d10f8g7h6i3j1k}\tabularnewline
{\footnotesize{}Ne} & {\footnotesize{}29s29p27d26f25g15h8i4j1k} & {\footnotesize{}13s12p12d10f8g7h6i3j1k} & {\footnotesize{}28s27p24d19f16g12h8i4j1k} & {\footnotesize{}13s12p11d10f8g7h6i3j1k}\tabularnewline
{\footnotesize{}Na} & {\footnotesize{}33s34p31d31f31g17h8i4j1k} & {\footnotesize{}16s13p13d10f9g7h6i3j1k} & {\footnotesize{}32s32p27d18f13g12h8i4j1k} & {\footnotesize{}15s13p13d10f9g7h6i3j1k}\tabularnewline
{\footnotesize{}Mg} & {\footnotesize{}32s33p30d30f30g16h8i4j1k} & {\footnotesize{}15s13p12d9f8g6h6i3j1k} & {\footnotesize{}31s31p26d18f13g11h8i4j1k} & {\footnotesize{}15s13p12d9f8g6h6i3j1k}\tabularnewline
{\footnotesize{}Al} & {\footnotesize{}33s33p33d32f31g18h8i4j1k} & {\footnotesize{}15s13p12d10f9g7h6i3j1k} & {\footnotesize{}32s32p27d19f14g12h8i4j1k} & {\footnotesize{}15s13p12d10f9g7h6i3j1k}\tabularnewline
{\footnotesize{}Si} & {\footnotesize{}32s33p32d32f32g18h8i4j1k} & {\footnotesize{}15s13p12d10f9g7h6i3j1k} & {\footnotesize{}31s31p27d20f16g12h8i4j1k} & {\footnotesize{}15s13p12d10f9g7h6i3j1k}\tabularnewline
{\footnotesize{}P} & {\footnotesize{}32s34p31d31f32g18h8i4j1k} & {\footnotesize{}15s13p12d10f9g7h6i3j1k} & {\footnotesize{}31s31p27d20f15g12h8i4j1k} & {\footnotesize{}15s13p12d10f9g7h6i3j1k}\tabularnewline
{\footnotesize{}S} & {\footnotesize{}32s33p31d31f32g18h8i4j1k} & {\footnotesize{}15s13p12d10f9g7h6i3j1k} & {\footnotesize{}31s31p27d19f15g13h8i4j1k} & {\footnotesize{}15s13p12d10f9g7h6i3j1k}\tabularnewline
{\footnotesize{}Cl} & {\footnotesize{}32s33p31d31f32g18h8i4j1k} & {\footnotesize{}15s13p12d10f9g7h6i3j1k} & {\footnotesize{}32s31p27d20f15g13h8i4j1k} & {\footnotesize{}15s13p12d10f9g7h6i3j1k}\tabularnewline
{\footnotesize{}Ar} & {\footnotesize{}32s33p31d31f31g18h8i4j1k} & {\footnotesize{}15s14p13d11f9g7h6i3j1k} & {\footnotesize{}31s31p27d19f15g12h8i4j1k} & {\footnotesize{}15s14p13d11f9g7h6i3j1k}\tabularnewline
\end{tabular}
\par\end{centering}
}

\caption{Compositions of contracted ABSs for 3ZaPa-NR and 4ZaPa-NR OBSs with
$\epsilon=10^{-5}$. The labeling of the angular momentum follows
the Gaussian'94 convention, which does not skip the letter $J$ for
$l=7$ unlike most other conventions. \label{tab:Compositions-of-contracted}}

\end{table*}

\section{Summary and Discussion \label{sec:Summary-and-Discussion}}

We have described an automated scheme for contracting auxiliary basis
sets (ABSs) for a given orbital basis set (OBS). The scheme works
with any type of atomic basis functions: in addition to the Gaussian-type
orbitals (GTOs) considered in this work, the scheme can also be straightforwardly
applied to other types of atomic basis functions, such as Slater-type
orbitals (STOs) and numerical atomic orbitals (NAOs). Our procedure
starts from a ``full'' ABS generated using the algorithm of \citeref{Lehtola2021_JCTC_6886}.
This ABS is then contracted and pruned of high-angular momentum (HAM)
functions to produce computationally efficient ABSs. 

By default, the scheme of \citeref{Lehtola2021_JCTC_6886} employs
a pivoted Cholesky decomposition of the electron repulsion integral
(ERI) tensor to choose the subset of orbital products from which the
auxiliary basis functions are chosen with another pivoted Cholesky
decomposition. We found in \citeref{Lehtola2021_JCTC_6886} that prescreening
the orbital products with a pivoted Cholesky of the ERIs significantly
decreases the number of primitive GTOs in the resulting autogenerated
auxiliary basis set, leading to better computational efficiency while
maintaining the same level of accuracy.

Interestingly, we find that the contracted auxiliary basis set obtained
with the singular value decomposition (SVD) of \ref{eq:W} comes out
similar even if the initial pivoted Cholesky decomposition of the
ERIs is not performed; that is, when the full set of orbital products
is employed to generate the auxiliary functions. The present scheme
thus appears to reproduce a similar number of contracted auxiliary
functions regardless of whether any initial screening of orbital products
is done.

Although GTOs incur additional cost from contractions, as their integrals
are usually evaluated in terms of Gaussian primitives, if NAOs are
used, the present scheme becomes especially interesting. As NAO integrals
are evaluated by quadrature, the linear transformation of the input
auxiliary functions can be simply carried out to their expansion coefficients;
the initialization with the pivoted Cholesky of the ERIs to choose
the product functions is therefore not needed in the context of NAO
calculations. We therefore believe that in addition to the application
of the present method to GTO basis sets presented in this work, our
scheme will be found to be extremely useful for NAO based calculations,
as well. We have recently developed modern software for atomic electronic
structure calculations employing high-order numerical methods that
allow rapid convergence to the complete basis set (CBS) limit with
few numerical basis functions,\citep{Lehtola2019_IJQC_25945,Lehtola2019_IJQC_25968,Lehtola2020_PRA_12516,Lehtola2023_JCTC_2502,Lehtola2023_JPCA_4180}
and hope to apply these techniques to NAO based calculations in the
near future, where the schemes of \citeref{Lehtola2021_JCTC_6886}
and this work will be of critical importance.
\begin{acknowledgement}
We thank the Academy of Finland for financial support under project
numbers 350282 and 353749. We thank CSC -- IT Centre for Science
(Espoo, Finland) for computational resources.
\end{acknowledgement}

\section*{Appendix: Conversion of Contraction Coefficients}

From the eigendecomposition, one obtains eigenvectors in terms of
Coulomb normalized primitives $\sum_{A}C_{A}^{\text{C}}N_{A}^{\text{C}}\phi_{A}(\boldsymbol{r})$,
while basis set input is handled in typical programs in terms of overlap
normalized primitives $\sum_{A}C_{A}^{\text{o}}N_{A}^{\text{o}}\phi_{A}(\boldsymbol{r})$.
Denoting the two normalization coefficients as $N_{A}^{\text{C}}=(A|A)^{-1/2}$
and $N_{A}^{\text{o}}=\langle A|A\rangle^{-1/2}$, where the overlap
of two unnormalized Gaussian primitives is\citep{Lehtola2020_JCP_134108}
\begin{equation}
\langle A|B\rangle=\frac{1}{2}\Gamma\left(l+\frac{3}{2}\right)(\alpha_{A}+\alpha_{B})^{-l-3/2}\delta_{ll_{A}}\delta_{ll_{B}}\label{eq:overlap}
\end{equation}
and the Coulomb overlap is given by\citep{Lehtola2021_JCTC_6886}
\begin{equation}
(A|B)=\frac{1}{4}\Gamma\left(l+\frac{3}{2}\right)\frac{(\alpha_{A}+\alpha_{B})^{-l-1/2}}{\alpha_{A}\alpha_{B}}\delta_{ll_{A}}\delta_{ll_{B}},\label{eq:coulomb-overlap}
\end{equation}
one finds the ratio
\begin{equation}
(A|B)=\frac{\alpha_{A}+\alpha_{B}}{2\alpha_{A}\alpha_{B}}\langle A|B\rangle\label{eq:ratio}
\end{equation}
from which the coefficient follows as
\[
C_{A}^{\text{o}}=\frac{N_{A}^{\text{o}}}{N_{A}^{\text{C}}}C_{A}^{\text{C}}=\sqrt{\frac{(A|A)}{\langle A|A\rangle}}C_{A}^{\text{C}}=\alpha_{A}^{1/2}C_{A}^{\text{C}},
\]
that is, the coefficients are multiplied by the square root of the
corresponding exponent.
\begin{tocentry}
\includegraphics[width=3.25in,height=1.75in,keepaspectratio]{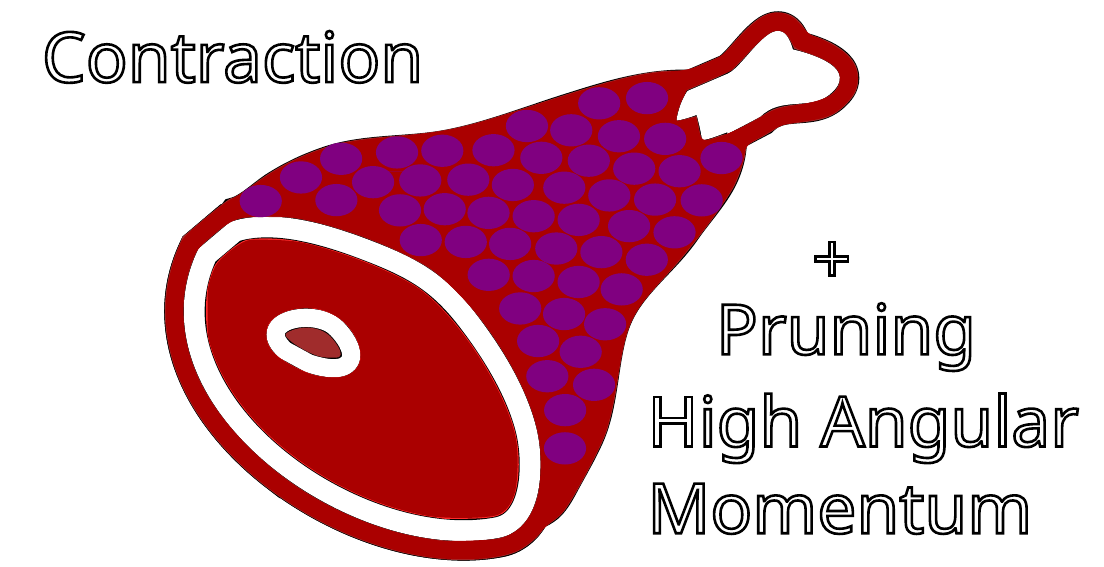}
\end{tocentry}
\bibliography{citations}

\providecommand{\latin}[1]{#1}
\makeatletter
\providecommand{\doi}
  {\begingroup\let\do\@makeother\dospecials
  \catcode`\{=1 \catcode`\}=2 \doi@aux}
\providecommand{\doi@aux}[1]{\endgroup\texttt{#1}}
\makeatother
\providecommand*\mcitethebibliography{\thebibliography}
\csname @ifundefined\endcsname{endmcitethebibliography}
  {\let\endmcitethebibliography\endthebibliography}{}
\begin{mcitethebibliography}{40}
\providecommand*\natexlab[1]{#1}
\providecommand*\mciteSetBstSublistMode[1]{}
\providecommand*\mciteSetBstMaxWidthForm[2]{}
\providecommand*\mciteBstWouldAddEndPuncttrue
  {\def\EndOfBibitem{\unskip.}}
\providecommand*\mciteBstWouldAddEndPunctfalse
  {\let\EndOfBibitem\relax}
\providecommand*\mciteSetBstMidEndSepPunct[3]{}
\providecommand*\mciteSetBstSublistLabelBeginEnd[3]{}
\providecommand*\EndOfBibitem{}
\mciteSetBstSublistMode{f}
\mciteSetBstMaxWidthForm{subitem}{(\alph{mcitesubitemcount})}
\mciteSetBstSublistLabelBeginEnd
  {\mcitemaxwidthsubitemform\space}
  {\relax}
  {\relax}

\bibitem[Whitten(1973)]{Whitten1973_JCP_4496}
Whitten,~J.~L. {Coulombic potential energy integrals and approximations}.
  \emph{J. Chem. Phys.} \textbf{1973}, \emph{58}, 4496\relax
\mciteBstWouldAddEndPuncttrue
\mciteSetBstMidEndSepPunct{\mcitedefaultmidpunct}
{\mcitedefaultendpunct}{\mcitedefaultseppunct}\relax
\EndOfBibitem
\bibitem[Baerends \latin{et~al.}(1973)Baerends, Ellis, and
  Ros]{Baerends1973_CP_41}
Baerends,~E.~J.; Ellis,~D.~E.; Ros,~P. {Self-consistent molecular
  Hartree--Fock--Slater calculations I. The computational procedure}.
  \emph{Chem. Phys.} \textbf{1973}, \emph{2}, 41--51\relax
\mciteBstWouldAddEndPuncttrue
\mciteSetBstMidEndSepPunct{\mcitedefaultmidpunct}
{\mcitedefaultendpunct}{\mcitedefaultseppunct}\relax
\EndOfBibitem
\bibitem[Dunlap \latin{et~al.}(1977)Dunlap, Connolly, and
  Sabin]{Dunlap1977_IJQC_81}
Dunlap,~B.~I.; Connolly,~J. W.~D.; Sabin,~J.~R. {On the applicability of
  LCAO-X$\alpha$ methods to molecules containing transition metal atoms: The
  nickel atom and nickel hydride}. \emph{Int. J. Quantum Chem.} \textbf{1977},
  \emph{12}, 81--87\relax
\mciteBstWouldAddEndPuncttrue
\mciteSetBstMidEndSepPunct{\mcitedefaultmidpunct}
{\mcitedefaultendpunct}{\mcitedefaultseppunct}\relax
\EndOfBibitem
\bibitem[Dunlap \latin{et~al.}(1979)Dunlap, Connolly, and
  Sabin]{Dunlap1979_JCP_3396}
Dunlap,~B.~I.; Connolly,~J. W.~D.; Sabin,~J.~R. {On some approximations in
  applications of X$\alpha$ theory}. \emph{J. Chem. Phys.} \textbf{1979},
  \emph{71}, 3396\relax
\mciteBstWouldAddEndPuncttrue
\mciteSetBstMidEndSepPunct{\mcitedefaultmidpunct}
{\mcitedefaultendpunct}{\mcitedefaultseppunct}\relax
\EndOfBibitem
\bibitem[Dunlap \latin{et~al.}(2010)Dunlap, R{\"{o}}sch, and
  Trickey]{Dunlap2010_MP_3167}
Dunlap,~B.~I.; R{\"{o}}sch,~N.; Trickey,~S.~B. {Variational fitting methods for
  electronic structure calculations}. \emph{Mol. Phys.} \textbf{2010},
  \emph{108}, 3167--3180\relax
\mciteBstWouldAddEndPuncttrue
\mciteSetBstMidEndSepPunct{\mcitedefaultmidpunct}
{\mcitedefaultendpunct}{\mcitedefaultseppunct}\relax
\EndOfBibitem
\bibitem[Vahtras \latin{et~al.}(1993)Vahtras, Alml{\"{o}}f, and
  Feyereisen]{Vahtras1993_CPL_514}
Vahtras,~O.; Alml{\"{o}}f,~J.; Feyereisen,~M.~W. {Integral approximations for
  LCAO-SCF calculations}. \emph{Chem. Phys. Lett.} \textbf{1993}, \emph{213},
  514--518\relax
\mciteBstWouldAddEndPuncttrue
\mciteSetBstMidEndSepPunct{\mcitedefaultmidpunct}
{\mcitedefaultendpunct}{\mcitedefaultseppunct}\relax
\EndOfBibitem
\bibitem[Smith \latin{et~al.}(2020)Smith, Burns, Simmonett, Parrish, Schieber,
  Galvelis, Kraus, Kruse, {Di Remigio}, Alenaizan, James, Lehtola, Misiewicz,
  Scheurer, Shaw, Schriber, Xie, Glick, Sirianni, O'Brien, Waldrop, Kumar,
  Hohenstein, Pritchard, Brooks, Schaefer, Sokolov, Patkowski, DePrince,
  Bozkaya, King, Evangelista, Turney, Crawford, and
  Sherrill]{Smith2020_JCP_184108}
Smith,~D. G.~A.; Burns,~L.~A.; Simmonett,~A.~C.; Parrish,~R.~M.;
  Schieber,~M.~C.; Galvelis,~R.; Kraus,~P.; Kruse,~H.; {Di Remigio},~R.;
  Alenaizan,~A.; James,~A.~M.; Lehtola,~S.; Misiewicz,~J.~P.; Scheurer,~M.;
  Shaw,~R.~A.; Schriber,~J.~B.; Xie,~Y.; Glick,~Z.~L.; Sirianni,~D.~A.;
  O'Brien,~J.~S.; Waldrop,~J.~M.; Kumar,~A.; Hohenstein,~E.~G.;
  Pritchard,~B.~P.; Brooks,~B.~R.; Schaefer,~H.~F.; Sokolov,~A.~Y.;
  Patkowski,~K.; DePrince,~A.~E.; Bozkaya,~U.; King,~R.~A.; Evangelista,~F.~A.;
  Turney,~J.~M.; Crawford,~T.~D.; Sherrill,~C.~D. {\sc Psi4} 1.4: Open-source
  software for high-throughput quantum chemistry. \emph{J. Chem. Phys.}
  \textbf{2020}, \emph{152}, 184108\relax
\mciteBstWouldAddEndPuncttrue
\mciteSetBstMidEndSepPunct{\mcitedefaultmidpunct}
{\mcitedefaultendpunct}{\mcitedefaultseppunct}\relax
\EndOfBibitem
\bibitem[Neese(2022)]{Neese2022_WCMS_1606}
Neese,~F. Software update: The ORCA program system---Version 5.0. \emph{WIREs
  Comput. Mol. Sci.} \textbf{2022}, \emph{12}, e1606\relax
\mciteBstWouldAddEndPuncttrue
\mciteSetBstMidEndSepPunct{\mcitedefaultmidpunct}
{\mcitedefaultendpunct}{\mcitedefaultseppunct}\relax
\EndOfBibitem
\bibitem[Shiozaki(2018)]{Shiozaki2018_WIRCMS_1331}
Shiozaki,~T. {BAGEL}: Brilliantly Advanced General Electronic-structure
  Library. \emph{Wiley Interdiscip. Rev. Comput. Mol. Sci.} \textbf{2018},
  \emph{8}, e1331\relax
\mciteBstWouldAddEndPuncttrue
\mciteSetBstMidEndSepPunct{\mcitedefaultmidpunct}
{\mcitedefaultendpunct}{\mcitedefaultseppunct}\relax
\EndOfBibitem
\bibitem[Geudtner \latin{et~al.}(2012)Geudtner, Calaminici,
  Carmona-Esp{\'{i}}ndola, del Campo, Dom{\'{i}}nguez-Soria, Moreno, Gamboa,
  Goursot, K{\"{o}}ster, Reveles, Mineva, V{\'{a}}squez-P{\'{e}}rez, Vela,
  Z{\'{u}}{\~{n}}inga-Gutierrez, and Salahub]{Geudtner2012_WIRCMS_548}
Geudtner,~G.; Calaminici,~P.; Carmona-Esp{\'{i}}ndola,~J.; del Campo,~J.~M.;
  Dom{\'{i}}nguez-Soria,~V.~D.; Moreno,~R.~F.; Gamboa,~G.~U.; Goursot,~A.;
  K{\"{o}}ster,~A.~M.; Reveles,~J.~U.; Mineva,~T.;
  V{\'{a}}squez-P{\'{e}}rez,~J.~M.; Vela,~A.;
  Z{\'{u}}{\~{n}}inga-Gutierrez,~B.; Salahub,~D.~R. {deMon2k}. \emph{Wiley
  Interdiscip. Rev. Comput. Mol. Sci.} \textbf{2012}, \emph{2}, 548--555\relax
\mciteBstWouldAddEndPuncttrue
\mciteSetBstMidEndSepPunct{\mcitedefaultmidpunct}
{\mcitedefaultendpunct}{\mcitedefaultseppunct}\relax
\EndOfBibitem
\bibitem[Hill(2013)]{Hill2013_IJQC_21}
Hill,~J.~G. {Gaussian basis sets for molecular applications}. \emph{Int. J.
  Quantum Chem.} \textbf{2013}, \emph{113}, 21--34\relax
\mciteBstWouldAddEndPuncttrue
\mciteSetBstMidEndSepPunct{\mcitedefaultmidpunct}
{\mcitedefaultendpunct}{\mcitedefaultseppunct}\relax
\EndOfBibitem
\bibitem[Pedersen \latin{et~al.}(2023)Pedersen, Lehtola, Galv{\'{a}}n, and
  Lindh]{Pedersen2023__}
Pedersen,~T.~B.; Lehtola,~S.; Galv{\'{a}}n,~I.~F.; Lindh,~R. The versatility of
  Cholesky decomposition in electronic structure theory. \textbf{2023}, \relax
\mciteBstWouldAddEndPunctfalse
\mciteSetBstMidEndSepPunct{\mcitedefaultmidpunct}
{}{\mcitedefaultseppunct}\relax
\EndOfBibitem
\bibitem[Aquilante \latin{et~al.}(2007)Aquilante, Lindh, and
  Pedersen]{Aquilante2007_JCP_114107}
Aquilante,~F.; Lindh,~R.; Pedersen,~T.~B. {Unbiased auxiliary basis sets for
  accurate two-electron integral approximations.} \emph{J. Chem. Phys.}
  \textbf{2007}, \emph{127}, 114107\relax
\mciteBstWouldAddEndPuncttrue
\mciteSetBstMidEndSepPunct{\mcitedefaultmidpunct}
{\mcitedefaultendpunct}{\mcitedefaultseppunct}\relax
\EndOfBibitem
\bibitem[Yang \latin{et~al.}(2007)Yang, Rendell, and
  Frisch]{Yang2007_JCP_74102}
Yang,~R.; Rendell,~A.~P.; Frisch,~M.~J. {Automatically generated Coulomb
  fitting basis sets: design and accuracy for systems containing H to Kr.}
  \emph{J. Chem. Phys.} \textbf{2007}, \emph{127}, 074102\relax
\mciteBstWouldAddEndPuncttrue
\mciteSetBstMidEndSepPunct{\mcitedefaultmidpunct}
{\mcitedefaultendpunct}{\mcitedefaultseppunct}\relax
\EndOfBibitem
\bibitem[Aquilante \latin{et~al.}(2009)Aquilante, Gagliardi, Pedersen, and
  Lindh]{Aquilante2009_JCP_154107}
Aquilante,~F.; Gagliardi,~L.; Pedersen,~T.~B.; Lindh,~R. {Atomic Cholesky
  decompositions: a route to unbiased auxiliary basis sets for density fitting
  approximation with tunable accuracy and efficiency.} \emph{J. Chem. Phys.}
  \textbf{2009}, \emph{130}, 154107\relax
\mciteBstWouldAddEndPuncttrue
\mciteSetBstMidEndSepPunct{\mcitedefaultmidpunct}
{\mcitedefaultendpunct}{\mcitedefaultseppunct}\relax
\EndOfBibitem
\bibitem[Ren \latin{et~al.}(2012)Ren, Rinke, Blum, Wieferink, Tkatchenko,
  Sanfilippo, Reuter, and Scheffler]{Ren2012_NJP_53020}
Ren,~X.; Rinke,~P.; Blum,~V.; Wieferink,~J.; Tkatchenko,~A.; Sanfilippo,~A.;
  Reuter,~K.; Scheffler,~M. {Resolution-of-identity approach to Hartree--Fock,
  hybrid density functionals, RPA, MP2 and GW with numeric atom-centered
  orbital basis functions}. \emph{New J. Phys.} \textbf{2012}, \emph{14},
  053020\relax
\mciteBstWouldAddEndPuncttrue
\mciteSetBstMidEndSepPunct{\mcitedefaultmidpunct}
{\mcitedefaultendpunct}{\mcitedefaultseppunct}\relax
\EndOfBibitem
\bibitem[Stoychev \latin{et~al.}(2017)Stoychev, Auer, and
  Neese]{Stoychev2017_JCTC_554}
Stoychev,~G.~L.; Auer,~A.~A.; Neese,~F. Automatic Generation of Auxiliary Basis
  Sets. \emph{J. Chem. Theory Comput.} \textbf{2017}, \emph{13}, 554--562\relax
\mciteBstWouldAddEndPuncttrue
\mciteSetBstMidEndSepPunct{\mcitedefaultmidpunct}
{\mcitedefaultendpunct}{\mcitedefaultseppunct}\relax
\EndOfBibitem
\bibitem[Lehtola(2021)]{Lehtola2021_JCTC_6886}
Lehtola,~S. Straightforward and Accurate Automatic Auxiliary Basis Set
  Generation for Molecular Calculations with Atomic Orbital Basis Sets.
  \emph{J. Chem. Theory Comput.} \textbf{2021}, \emph{17}, 6886--6900\relax
\mciteBstWouldAddEndPuncttrue
\mciteSetBstMidEndSepPunct{\mcitedefaultmidpunct}
{\mcitedefaultendpunct}{\mcitedefaultseppunct}\relax
\EndOfBibitem
\bibitem[Beebe and Linderberg(1977)Beebe, and Linderberg]{Beebe1977_IJQC_683}
Beebe,~N. H.~F.; Linderberg,~J. {Simplifications in the Two-Electron Integral
  Array in Molecular Calculations}. \emph{Int. J. Quant. Chem.} \textbf{1977},
  \emph{12}, 683--705\relax
\mciteBstWouldAddEndPuncttrue
\mciteSetBstMidEndSepPunct{\mcitedefaultmidpunct}
{\mcitedefaultendpunct}{\mcitedefaultseppunct}\relax
\EndOfBibitem
\bibitem[Harbrecht \latin{et~al.}(2012)Harbrecht, Peters, and
  Schneider]{Harbrecht2012_ANM_428}
Harbrecht,~H.; Peters,~M.; Schneider,~R. {On the low-rank approximation by the
  pivoted Cholesky decomposition}. \emph{Appl. Numer. Math.} \textbf{2012},
  \emph{62}, 428--440\relax
\mciteBstWouldAddEndPuncttrue
\mciteSetBstMidEndSepPunct{\mcitedefaultmidpunct}
{\mcitedefaultendpunct}{\mcitedefaultseppunct}\relax
\EndOfBibitem
\bibitem[Lehtola(2019)]{Lehtola2019_IJQC_25968}
Lehtola,~S. A review on non-relativistic, fully numerical electronic structure
  calculations on atoms and diatomic molecules. \emph{Int. J. Quantum Chem.}
  \textbf{2019}, \emph{119}, e25968\relax
\mciteBstWouldAddEndPuncttrue
\mciteSetBstMidEndSepPunct{\mcitedefaultmidpunct}
{\mcitedefaultendpunct}{\mcitedefaultseppunct}\relax
\EndOfBibitem
\bibitem[Lehtola(2019)]{Lehtola2019_IJQC_25945}
Lehtola,~S. Fully numerical {Hartree}--{Fock} and density functional
  calculations. {I}. {Atoms}. \emph{Int. J. Quantum Chem.} \textbf{2019},
  \emph{119}, e25945\relax
\mciteBstWouldAddEndPuncttrue
\mciteSetBstMidEndSepPunct{\mcitedefaultmidpunct}
{\mcitedefaultendpunct}{\mcitedefaultseppunct}\relax
\EndOfBibitem
\bibitem[Roothaan(1951)]{Roothaan1951_RMP_69}
Roothaan,~C. C.~J. New Developments in Molecular Orbital Theory. \emph{Rev.
  Mod. Phys.} \textbf{1951}, \emph{23}, 69--89\relax
\mciteBstWouldAddEndPuncttrue
\mciteSetBstMidEndSepPunct{\mcitedefaultmidpunct}
{\mcitedefaultendpunct}{\mcitedefaultseppunct}\relax
\EndOfBibitem
\bibitem[Lehtola(2019)]{Lehtola2019_JCP_241102}
Lehtola,~S. Curing basis set overcompleteness with pivoted {Cholesky}
  decompositions. \emph{J. Chem. Phys.} \textbf{2019}, \emph{151}, 241102\relax
\mciteBstWouldAddEndPuncttrue
\mciteSetBstMidEndSepPunct{\mcitedefaultmidpunct}
{\mcitedefaultendpunct}{\mcitedefaultseppunct}\relax
\EndOfBibitem
\bibitem[Lehtola(2020)]{Lehtola2020_PRA_32504}
Lehtola,~S. Accurate reproduction of strongly repulsive interatomic potentials.
  \emph{Phys. Rev. A} \textbf{2020}, \emph{101}, 032504\relax
\mciteBstWouldAddEndPuncttrue
\mciteSetBstMidEndSepPunct{\mcitedefaultmidpunct}
{\mcitedefaultendpunct}{\mcitedefaultseppunct}\relax
\EndOfBibitem
\bibitem[K{\'{a}}llay(2014)]{Kallay2014_JCP_244113}
K{\'{a}}llay,~M. {A systematic way for the cost reduction of density fitting
  methods}. \emph{J. Chem. Phys.} \textbf{2014}, \emph{141}, 244113\relax
\mciteBstWouldAddEndPuncttrue
\mciteSetBstMidEndSepPunct{\mcitedefaultmidpunct}
{\mcitedefaultendpunct}{\mcitedefaultseppunct}\relax
\EndOfBibitem
\bibitem[Weigend(2008)]{Weigend2008_JCC_167}
Weigend,~F. {Hartree--Fock exchange fitting basis sets for H to Rn}. \emph{J.
  Comput. Chem.} \textbf{2008}, \emph{29}, 167--175\relax
\mciteBstWouldAddEndPuncttrue
\mciteSetBstMidEndSepPunct{\mcitedefaultmidpunct}
{\mcitedefaultendpunct}{\mcitedefaultseppunct}\relax
\EndOfBibitem
\bibitem[Manni \latin{et~al.}(2023)Manni, Galv{\'{a}}n, Alavi, Aleotti,
  Aquilante, Autschbach, Avagliano, Baiardi, Bao, Battaglia, Birnoschi,
  Blanco-Gonz{\'{a}}lez, Bokarev, Broer, Cacciari, Calio, Carlson, Couto,
  Cerd{\'{a}}n, Chibotaru, Chilton, Church, Conti, Coriani,
  Cu{\'{e}}llar-Zuquin, Daoud, Dattani, Decleva, de~Graaf, Delcey, Vico,
  Dobrautz, Dong, Feng, Ferr{\'{e}}, Filatov(Gulak), Gagliardi, Garavelli,
  Gonz{\'{a}}lez, Guan, Guo, Hennefarth, Hermes, Hoyer, Huix-Rotllant, Jaiswal,
  Kaiser, Kaliakin, Khamesian, King, Kochetov, Kro{\'{s}}nicki, Kumaar,
  Larsson, Lehtola, Lepetit, Lischka, R{\'{\i}}os, Lundberg, Ma, Mai,
  Marquetand, Merritt, Montorsi, Mörchen, Nenov, Nguyen, Nishimoto, Oakley,
  Olivucci, Oppel, Padula, Pandharkar, Phung, Plasser, Raggi, Rebolini, Reiher,
  Rivalta, Roca-Sanju{\'{a}}n, Romig, Safari, S{\'{a}}nchez-Mansilla, Sand,
  Schapiro, Scott, Segarra-Mart{\'{\i}}, Segatta, Sergentu, Sharma, Shepard,
  Shu, Staab, Straatsma, S{\o}rensen, Tenorio, Truhlar, Ungur, Vacher,
  Veryazov, Vo{\ss}, Weser, Wu, Yang, Yarkony, Zhou, Zobel, and
  Lindh]{Manni2023_JCTC_}
Manni,~G.~L.; Galv{\'{a}}n,~I.~F.; Alavi,~A.; Aleotti,~F.; Aquilante,~F.;
  Autschbach,~J.; Avagliano,~D.; Baiardi,~A.; Bao,~J.~J.; Battaglia,~S.;
  Birnoschi,~L.; Blanco-Gonz{\'{a}}lez,~A.; Bokarev,~S.~I.; Broer,~R.;
  Cacciari,~R.; Calio,~P.~B.; Carlson,~R.~K.; Couto,~R.~C.; Cerd{\'{a}}n,~L.;
  Chibotaru,~L.~F.; Chilton,~N.~F.; Church,~J.~R.; Conti,~I.; Coriani,~S.;
  Cu{\'{e}}llar-Zuquin,~J.; Daoud,~R.~E.; Dattani,~N.; Decleva,~P.;
  de~Graaf,~C.; Delcey,~M.~G.; Vico,~L.~D.; Dobrautz,~W.; Dong,~S.~S.;
  Feng,~R.; Ferr{\'{e}},~N.; Filatov(Gulak),~M.; Gagliardi,~L.; Garavelli,~M.;
  Gonz{\'{a}}lez,~L.; Guan,~Y.; Guo,~M.; Hennefarth,~M.~R.; Hermes,~M.~R.;
  Hoyer,~C.~E.; Huix-Rotllant,~M.; Jaiswal,~V.~K.; Kaiser,~A.; Kaliakin,~D.~S.;
  Khamesian,~M.; King,~D.~S.; Kochetov,~V.; Kro{\'{s}}nicki,~M.; Kumaar,~A.~A.;
  Larsson,~E.~D.; Lehtola,~S.; Lepetit,~M.-B.; Lischka,~H.; R{\'{\i}}os,~P.~L.;
  Lundberg,~M.; Ma,~D.; Mai,~S.; Marquetand,~P.; Merritt,~I. C.~D.;
  Montorsi,~F.; Mörchen,~M.; Nenov,~A.; Nguyen,~V. H.~A.; Nishimoto,~Y.;
  Oakley,~M.~S.; Olivucci,~M.; Oppel,~M.; Padula,~D.; Pandharkar,~R.;
  Phung,~Q.~M.; Plasser,~F.; Raggi,~G.; Rebolini,~E.; Reiher,~M.; Rivalta,~I.;
  Roca-Sanju{\'{a}}n,~D.; Romig,~T.; Safari,~A.~A.; S{\'{a}}nchez-Mansilla,~A.;
  Sand,~A.~M.; Schapiro,~I.; Scott,~T.~R.; Segarra-Mart{\'{\i}},~J.;
  Segatta,~F.; Sergentu,~D.-C.; Sharma,~P.; Shepard,~R.; Shu,~Y.; Staab,~J.~K.;
  Straatsma,~T.~P.; S{\o}rensen,~L.~K.; Tenorio,~B. N.~C.; Truhlar,~D.~G.;
  Ungur,~L.; Vacher,~M.; Veryazov,~V.; Vo{\ss},~T.~A.; Weser,~O.; Wu,~D.;
  Yang,~X.; Yarkony,~D.; Zhou,~C.; Zobel,~J.~P.; Lindh,~R. The {OpenMolcas}
  \emph{Web}: A Community-Driven Approach to Advancing Computational Chemistry.
  \emph{J. Chem. Theory Comput.} \textbf{2023}, \relax
\mciteBstWouldAddEndPunctfalse
\mciteSetBstMidEndSepPunct{\mcitedefaultmidpunct}
{}{\mcitedefaultseppunct}\relax
\EndOfBibitem
\bibitem[Ranasinghe and Petersson(2013)Ranasinghe, and
  Petersson]{Ranasinghe2013_JCP_144104}
Ranasinghe,~D.~S.; Petersson,~G.~A. {CCSD(T)/CBS} atomic and molecular
  benchmarks for {H} through {Ar}. \emph{J. Chem. Phys.} \textbf{2013},
  \emph{138}, 144104\relax
\mciteBstWouldAddEndPuncttrue
\mciteSetBstMidEndSepPunct{\mcitedefaultmidpunct}
{\mcitedefaultendpunct}{\mcitedefaultseppunct}\relax
\EndOfBibitem
\bibitem[Pritchard \latin{et~al.}(2019)Pritchard, Altarawy, Didier, Gibson, and
  Windus]{Pritchard2019_JCIM_4814}
Pritchard,~B.~P.; Altarawy,~D.; Didier,~B.; Gibson,~T.~D.; Windus,~T.~L. New
  {Basis} {Set} {Exchange}: An Open, Up-to-Date Resource for the Molecular
  Sciences Community. \emph{J. Chem. Inf. Model.} \textbf{2019}, \emph{59},
  4814--4820\relax
\mciteBstWouldAddEndPuncttrue
\mciteSetBstMidEndSepPunct{\mcitedefaultmidpunct}
{\mcitedefaultendpunct}{\mcitedefaultseppunct}\relax
\EndOfBibitem
\bibitem[Lehtola \latin{et~al.}(2012)Lehtola, Hakala, Sakko, and
  H{\"{a}}m{\"{a}}l{\"{a}}inen]{Lehtola2012_JCC_1572}
Lehtola,~J.; Hakala,~M.; Sakko,~A.; H{\"{a}}m{\"{a}}l{\"{a}}inen,~K. {ERKALE}
  -- A flexible program package for X-ray properties of atoms and molecules.
  \emph{J. Comput. Chem.} \textbf{2012}, \emph{33}, 1572--1585\relax
\mciteBstWouldAddEndPuncttrue
\mciteSetBstMidEndSepPunct{\mcitedefaultmidpunct}
{\mcitedefaultendpunct}{\mcitedefaultseppunct}\relax
\EndOfBibitem
\bibitem[Lehtola(2023)]{Lehtola2018__a}
Lehtola,~S. {ERKALE -- HF/DFT from Hel}. 2023;
  \url{https://github.com/susilehtola/erkale}, Accessed 23 March 2023.\relax
\mciteBstWouldAddEndPunctfalse
\mciteSetBstMidEndSepPunct{\mcitedefaultmidpunct}
{}{\mcitedefaultseppunct}\relax
\EndOfBibitem
\bibitem[Karton \latin{et~al.}(2017)Karton, Sylvetsky, and
  Martin]{Karton2017_JCC_2063}
Karton,~A.; Sylvetsky,~N.; Martin,~J. M.~L. {W4-17: A diverse and
  high-confidence dataset of atomization energies for benchmarking high-level
  electronic structure methods}. \emph{J. Comput. Chem.} \textbf{2017},
  \emph{38}, 2063--2075\relax
\mciteBstWouldAddEndPuncttrue
\mciteSetBstMidEndSepPunct{\mcitedefaultmidpunct}
{\mcitedefaultendpunct}{\mcitedefaultseppunct}\relax
\EndOfBibitem
\bibitem[Weigend and Ahlrichs(2005)Weigend, and Ahlrichs]{Weigend2005_PCCP_305}
Weigend,~F.; Ahlrichs,~R. {Balanced basis sets of split valence, triple zeta
  valence and quadruple zeta valence quality for H to Rn: Design and assessment
  of accuracy.} \emph{Phys. Chem. Chem. Phys.} \textbf{2005}, \emph{7},
  3297--305\relax
\mciteBstWouldAddEndPuncttrue
\mciteSetBstMidEndSepPunct{\mcitedefaultmidpunct}
{\mcitedefaultendpunct}{\mcitedefaultseppunct}\relax
\EndOfBibitem
\bibitem[Frisch \latin{et~al.}(2016)Frisch, Trucks, Schlegel, Scuseria, Robb,
  Cheeseman, Scalmani, Barone, Petersson, Nakatsuji, Li, Caricato, Marenich,
  Bloino, Janesko, Gomperts, Mennucci, Hratchian, Ortiz, Izmaylov, Sonnenberg,
  Williams-Young, Ding, Lipparini, Egidi, Goings, Peng, Petrone, Henderson,
  Ranasinghe, Zakrzewski, Gao, Rega, Zheng, Liang, Hada, Ehara, Toyota, Fukuda,
  Hasegawa, Ishida, Nakajima, Honda, Kitao, Nakai, Vreven, Throssell,
  {Montgomery Jr.}, Peralta, Ogliaro, Bearpark, Heyd, Brothers, Kudin,
  Staroverov, Keith, Kobayashi, Normand, Raghavachari, Rendell, Burant,
  Iyengar, Tomasi, Cossi, Millam, Klene, Adamo, Cammi, Ochterski, Martin,
  Morokuma, Farkas, Foresman, and Fox]{Frisch2016__}
Frisch,~M.~J.; Trucks,~G.~W.; Schlegel,~H.~B.; Scuseria,~G.~E.; Robb,~M.~A.;
  Cheeseman,~J.~R.; Scalmani,~G.; Barone,~V.; Petersson,~G.~A.; Nakatsuji,~H.;
  Li,~X.; Caricato,~M.; Marenich,~A.~V.; Bloino,~J.; Janesko,~B.~G.;
  Gomperts,~R.; Mennucci,~B.; Hratchian,~H.~P.; Ortiz,~J.~V.; Izmaylov,~A.~F.;
  Sonnenberg,~J.~L.; Williams-Young,~D.; Ding,~F.; Lipparini,~F.; Egidi,~F.;
  Goings,~J.; Peng,~B.; Petrone,~A.; Henderson,~T.; Ranasinghe,~D.;
  Zakrzewski,~V.~G.; Gao,~J.; Rega,~N.; Zheng,~G.; Liang,~W.; Hada,~M.;
  Ehara,~M.; Toyota,~K.; Fukuda,~R.; Hasegawa,~J.; Ishida,~M.; Nakajima,~T.;
  Honda,~Y.; Kitao,~O.; Nakai,~H.; Vreven,~T.; Throssell,~K.; {Montgomery
  Jr.},~J.~A.; Peralta,~J.~E.; Ogliaro,~F.; Bearpark,~M.~J.; Heyd,~J.~J.;
  Brothers,~E.~N.; Kudin,~K.~N.; Staroverov,~V.~N.; Keith,~T.~A.;
  Kobayashi,~R.; Normand,~J.; Raghavachari,~K.; Rendell,~A.~P.; Burant,~J.~C.;
  Iyengar,~S.~S.; Tomasi,~J.; Cossi,~M.; Millam,~J.~M.; Klene,~M.; Adamo,~C.;
  Cammi,~R.; Ochterski,~J.~W.; Martin,~R.~L.; Morokuma,~K.; Farkas,~O.;
  Foresman,~J.~B.; Fox,~D.~J. {Gaussian 16 Revision B.01}. 2016\relax
\mciteBstWouldAddEndPuncttrue
\mciteSetBstMidEndSepPunct{\mcitedefaultmidpunct}
{\mcitedefaultendpunct}{\mcitedefaultseppunct}\relax
\EndOfBibitem
\bibitem[Lehtola(2020)]{Lehtola2020_PRA_12516}
Lehtola,~S. Fully numerical calculations on atoms with fractional occupations
  and range-separated exchange functionals. \emph{Phys. Rev. A} \textbf{2020},
  \emph{101}, 012516\relax
\mciteBstWouldAddEndPuncttrue
\mciteSetBstMidEndSepPunct{\mcitedefaultmidpunct}
{\mcitedefaultendpunct}{\mcitedefaultseppunct}\relax
\EndOfBibitem
\bibitem[Lehtola(2023)]{Lehtola2023_JCTC_2502}
Lehtola,~S. Meta-{GGA} Density Functional Calculations on Atoms with
  Spherically Symmetric Densities in the Finite Element Formalism. \emph{J.
  Chem. Theory Comput.} \textbf{2023}, \emph{19}, 2502--2517\relax
\mciteBstWouldAddEndPuncttrue
\mciteSetBstMidEndSepPunct{\mcitedefaultmidpunct}
{\mcitedefaultendpunct}{\mcitedefaultseppunct}\relax
\EndOfBibitem
\bibitem[Lehtola(2023)]{Lehtola2023_JPCA_4180}
Lehtola,~S. Atomic Electronic Structure Calculations with {Hermite}
  Interpolating Polynomials. \emph{J. Phys. Chem. A} \textbf{2023}, \emph{127},
  4180--4193\relax
\mciteBstWouldAddEndPuncttrue
\mciteSetBstMidEndSepPunct{\mcitedefaultmidpunct}
{\mcitedefaultendpunct}{\mcitedefaultseppunct}\relax
\EndOfBibitem
\bibitem[Lehtola(2020)]{Lehtola2020_JCP_134108}
Lehtola,~S. Polarized {Gaussian} basis sets from one-electron ions. \emph{J.
  Chem. Phys.} \textbf{2020}, \emph{152}, 134108\relax
\mciteBstWouldAddEndPuncttrue
\mciteSetBstMidEndSepPunct{\mcitedefaultmidpunct}
{\mcitedefaultendpunct}{\mcitedefaultseppunct}\relax
\EndOfBibitem
\end{mcitethebibliography}

\end{document}